\documentclass[12pt]{JHEP}
\usepackage{graphicx,amsmath,amssymb}
\usepackage{epsfig,multicol}
\usepackage{epsf}
\usepackage{epstopdf}
\input{epsf.sty}

\def\bx{{\bf x}}

\def\CL{{\cal L}}
\def\cO{{\cal O}}

\def\talpha{{\tilde \alpha}}
\def\tphi{{\tilde \phi}}
\def\tpsi{{\tilde \psi}}
\def\tH{{\tilde H}}

\def\m10{M_{10}}
\def\mpl{M_{\rm Pl}}
\def\half{\frac{1}{2}}

\newcommand{\be}{\begin{equation}}
\newcommand{\ee}{\end{equation}}
\newcommand{\bea}{\begin{eqnarray}}
\newcommand{\eea}{\end{eqnarray}}
\newcommand{\barr}{\begin{array}}
\newcommand{\earr}{\end{array}}

\newcommand{\ba}{\begin{eqnarray}}
\newcommand{\ea}{\end{eqnarray}}

\title{Heating in Brane Inflation and \\ Hidden Dark Matter}
\author{Xingang Chen$^{1,2}$ and S.-H. Henry Tye$^3$
\\ \small{\em $^1$Institute for Fundamental Theory \\
Department of Physics, University of Florida,
Gainesville, FL 32611}
\vskip .1cm
\\ \small{\em $^2$ISCAP, Physics Department, Columbia University \\
New York, NY 10027}
\vskip .1cm
\\\small{\em $^3$Newman Laboratory for Elementary Particle Physics \\
Cornell University, Ithaca, NY 14853} }

\abstract{Towards the end of brane inflation, the brane pair
annihilation produces massive closed strings. 
The transfer of this energy to Standard Model (SM) open string modes
depends on
where the SM branes and the brane annihilation are located:
in the bulk, in the same throat or in different throats. 
We find that, in
all cases as long as the brane annihilation and the SM branes are not
both in the bulk,
the transfer of energy to start the hot big bang epoch can be
efficient enough
to be compatible with big bang nucleosynthesis. The suppression of the
abundance of the graviton and its Kaluza-Klein (KK) thermal relics
follows from the warped geometry in flux compactification.
This works out even in the scenarios where a long period of
tunneling is expected.
In the multi-throat scenario, we find a dynamical mechnism of
selecting a long throat as the SM throat.
We establish three new dark matter candidates: KK modes with
specific angular momentum in the SM throat, those in the
brane annihilation throat, and different matters generated by
KK modes tunneled to other throats. Since the latter two
couple to the visible matter sector only through graviton mediation, 
they behave
as hidden dark matter. Hidden dark matter has novel implications on
the dark matter coincidence problem and the high energy cosmic rays.
}

\preprint{hep-th/0602136 \\ UFIFT-HEP-06-4 \\ CU-TP-1144}

\begin{document}

%\maketitle

%\setcounter{page}{0}
%\thispagestyle{empty}

%\baselineskip 16pt plus 2pt minus 2pt

%\baselineskip=18pt

\clearpage

\section{Introduction and summary}

\subsection{Introduction}
One of the most interesting aspects of brane
inflation \cite{Dvali:1998pa,Burgess:2001fx,Dvali:2001fw} is its close
connection to experiments. This includes the spectral index,
tensor mode, non-Gaussianities and cosmic strings --- ranging from
astronomical observations, gravitational wave detections to cosmic
microwave background radiation (CMBR) measurements.

Flux compactification in type IIB string theory has become an ideal
setup to realize the brane inflation. This is not only because it
provides promising mechanisms to stabilize
moduli \cite{Giddings:2001yu,Kachru:2003aw,Grana:2005jc} 
and therefore a consistent string theory setup, but also because the flux 
induced warped space has become a very interesting new ingredient for model
building. 

First, warped throats provide different mechanisms to achieve brane
inflation. Anti-D3-branes naturally settle down in throats. The warped geometry 
red-shifts the brane tension and reduces (easily by a large factor) the
mutual attraction between the branes and antibranes. 
This effect helps the slow-roll inflation, as in the 
KKLMMT scenario \cite{Kachru:2003sx,Firouzjahi:2005dh}. 
Since the inflaton (the position of a D3-brane)
is an open string mode, its kinetic term appears in the form of a DBI action.
In warped space this imposes a speed-limit to slow down the brane
motion in case
the radiatively corrected potential becomes too steep. 
This mechanism generates the DBI
inflation \cite{Silverstein:2003hf,Alishahiha:2004eh,Chen:2004gc,Chen:2005ad}.
Warped geometry together
with the DBI action essentially guarantees enough e-folds of inflation.
At the same time, effects of these two different mechanisms, and
its combinations \cite{Shandera:2006ax}, can give rise to interesting
observational signatures.

Second, the simple geometric picture allows us to explore variations
of the simplest scenario involving multiple throats and branes.
The multi-throat configuration arising from the flux
compactification provides a natural setup to generate candidate
inflatons \cite{Chen:2004gc}. 
Antibranes settle down in throats and annihilate against fluxes 
quantum mechanically \cite{Kachru:2002gs,DeWolfe:2004qx}. In 4-d view
point, this process can start the brane
inflationary epoch by eternally creating
bubbles in the old inflation background. Within each bubble, the
flux-antibrane annihilation naturally generate many D3-branes. The
vacuum energy density within the bubble can still be large. These
liberated D3-branes can roll out of the throat to other places and
generate different types of brane inflation, such as the slow-roll or DBI
inflation, whenever required conditions are satisfied. Different
bubbles can also correspond to annihilations in different throats --
the frequencies of which are determined by the relative lifetimes of
the antibranes. In some cases, multiple branes may also be used to
dynamically tune a flat potential \cite{Cline:2005ty}.

Third, hierarchical range of string scales in multi-throat
compactification provides different opportunities to observe stringy
physics. Low string scale throat can make strings relatively easier to
be observed in colliders \cite{Randall:1999ee}, or can imprint stringy
information on the density
perturbations at large scale
\cite{Chen:2005ad,Chen:2005fe}. High string scale throat
can store high tension cosmic strings at the end of brane inflation. The
allowed range of cosmic string tension overlaps with observational
abilities \cite{Jones:2002cv,Sarangi:2002yt,Copeland:2003bj,Polchinski:2004ia}.
In this paper, we propose another possible consequence of brane inflation, 
namely hidden dark matter.

The crucial step that links the inflationary epoch to the hot big bang epoch
is the heating at the end of the inflation. This is known as the graceful exit.
However, the heating in both the single- and multi-throat
configurations remain so far the least understood aspect of brane 
inflation. Namely, how can inflationary energy be efficiently
transfered to heat up the Standard Model particles, and be compatible
with the well-understood late-time cosmological evolution? 
This is the heating problem (also called reheating or preheating
problem). To see why this is quite a non-trivial question, we first
look at the end process of the brane inflation.

The brane inflation typically ends when branes annihilate
antibranes. Significant insights have been gained into such a process
recently \cite{Sen:2002nu,Sen:2002in,Sen:2004nf}. 
Tachyonic modes appear when the brane-antibrane distance approaches the string scale
and the annihilation process may be described by the tachyon rolling
\cite{Shiu:2002xp,Cline:2002it}.
No matter whether there are adjacent extra branes surviving
such an annihilation, the initial end product is expected to be dominated by
non-relativistic heavy closed
strings \cite{Lambert:2003zr,Chen:2003xq,Leblond:2005km}.
These will further be involved to 
create lighter closed strings, KK modes, gravitons and open
strings. (In this paper, we use KK modes as an abbreviation of the KK
modes of massless closed strings.) We know from observations that,
during big bang 
nucleosynthesis (BBN), the density of gravitons can be no more than
a few percent of the total energy density of the universe. The rest is
contributed by the Standard Model (SM) particles (mostly photons,
neutrinos and electrons), which are open strings attached to a stack
of SM (anti-)branes.
We also know that the density of any non-relativistic relics can be no
more than 10 times that of the baryons.
Therefore the question becomes how the brane annihilation products,
originally 
dominated by the closed string degrees of freedom, can eventually
become the required light open string degrees of freedom living on the
SM branes, with a negligible graviton density and a non-lethal amount
of stable relics.

Depending on where the SM branes and brane annihilation are, we may
consider 3 possibilities : \\
$\bullet$ when the SM branes are in the bulk (e.g., D7-branes wrapping
a 4-cycle), while the brane annihilation happens in a throat; or vice
versa; or they are both in the bulk; \\ 
$\bullet$ the single throat scenario, in which case the SM (anti-)branes are
in the same throat where the inflationary brane annihilation takes
place; note that the model
can still have other throats around; \\
$\bullet$  the multi-throat scenario, where the inflationary branes
annihilate in one throat (A-throat) while 
the SM branes are sitting in another throat (S-throat).

Both the slow-roll and DBI inflation can be realized in different
scenarios. Depending on whether one uses the throat to solve the
hierarchy problem or to stabilize the cosmic strings, different models
have different requirements. 
For DBI inflation, a single A-throat may both fit the density
perturbation and have a large hierarchy, because its prediction on the
density perturbations can be independent of the A-throat warp factor
\cite{Chen:2004gc,Chen:2005ad}.
In slow-roll brane inflationary models, the CMBR data typically
requires only a modestly warped
throat while solving the hierarchy problem requires substantially more 
warping, so the multi-throat scenario seems to be preferred as it can
easily
accommodate this very different warping properties.
In both inflation models, 
the single throat versus the
multi-throat scenario has another important implication. Brane annihilation is 
expected to produce all types of strings/defects that are not
forbidden. This includes
cosmic strings, which are simply massive large fundamental strings,
D-1 strings and/or
axionic strings. They tend to survive much better in a throat without branes.  
In the single throat scenario, due to their interaction with SM
branes, they are unlikely to survive cosmologically due to their
potential instabilities
\cite{Copeland:2003bj,Leblond:2004uc,Polchinski:2005bg,Blanco-Pillado:2005xx}. 
So the prediction of cosmic strings is much firmer in the multi-throat
scenario,
provided heating after inflation is not a problem. 
This calls for a careful examination of heating in the multi-throat scenario.

A number of studies have been done to address this heating problem
\cite{Barnaby:2004gg,Kofman:2005yz,Chialva:2005zy,Frey:2005jk}.
An important observation is that, because the KK mode wave function
is peaked at the bottom of the throat, its interaction with
particles located at the IR side is much enhanced compared to that
with the
graviton. This is essentially along the line of Randall-Sundrum (RS)
model \cite{Randall:1999ee}. Because of this,
the graviton emission branching ratio during the brane decay and KK
evolution is suppressed by powers of warp
factors \cite{Barnaby:2004gg,Kofman:2005yz,Chialva:2005zy}. Such a
suppression is absent if the brane annihilation happens in the bulk.

But several concerns have also been
raised \cite{Barnaby:2004gg,Kofman:2005yz,Chialva:2005zy}. 
Since the KK modes in
a throat can have (approximately) conserved angular momenta, it is
important to estimate its relic density. This was previously estimated
to be extremely large \cite{Kofman:2005yz}, 
and therefore any long-living KK relics seemed to be very
dangerous to the BBN. Serious problems have also been raised in the
multi-throat scenario where the question of heating become especially
sharper \cite{Barnaby:2004gg,Chialva:2005zy}. 
For example, in the multi-throat heating, the energy released
from the brane annihilation has to be transfered to anther throat by
tunneling, which is relatively a very long process. During this
period, it
seems that the KK modes in the A-throat is doing nothing
but annihilate to gravitons.
Requiring the initial annihilation rate to be smaller than the
tunneling rate imposes very restrictive conditions on parameters of the
model \cite{Chialva:2005zy}. If generalized to the warped
compactification where the tunneling rate is even much smaller, the
analogous condition would rule out the multi-throat
heating scenario.

However, a more detailed thermodynamic evolution of the heating
process, especially that of the KK particles, has not been carefully
studied. Such a thermal history is important for obtaining various
estimates and assessing 
the viability of the brane inflation heating. This
paper is a step towards this direction. 
Other new ingredients are, instead of the Randall-Sundrum (RS) 1-d warped
geometry, we consider a 6-d warped geometry more relevant to the brane inflation
in IIB theory (although the results can be easily evaluated and
generalized to the RS case). 
This will cause interesting modifications in various estimates on
tunneling rates and interaction rates.
In a realistic compactification, throats
are typically
separated in the bulk, which tends to generate resonance effects in
the tunneling from one throat to another. We expect the compactification 
volume to be dominated by the bulk, another important ingredient in the 
success of the graceful exit.

We will find many qualitatively different results compared to those in
Ref.\cite{Barnaby:2004gg,Kofman:2005yz,Chialva:2005zy}.
For example, the KK modes will decouple typically after they become
non-relativistic. Because of the warping enhanced KK
self-interactions, their relic density turn out to be suppressed by
powers of warp factors and the bulk size. In the multi-throat case,
the KK relic abundance is further suppressed due to an extra matter
dominated tunneling phase.
Also for the multi-throat case, although the
graviton production rate is initially much greater than the tunneling
rate, the rate gets diluted by the spatial expansion. So the graviton
production is significant only in the early epoch and its eventual abundance is
also suppressed by similar factors. The most stringent constraint on
the tunneling is just that it cannot be too long to over-cool the
universe.

So a final picture emerges is that a tower of KK particles gradually
lowers as the universe expands, sometimes intervened by a long period
of tunneling, and eventually transfers most of its entropy to the
Standard Model particles at the bottom of the spectrum. 
The KK relic and graviton density are both reasonably suppressed by
various factors due to the warped compactification.

During the course of this study, we will find many other interesting
phenomena arising from the multi-throat heating. First, there is a
dynamical process that selects a long throat to be heated. This is
because the dense spectrum in long throats makes the level
matching of the energy eigenstates, a necessary condition for
tunneling between throats, easier to satisfy. This may provide a
dynamical explanation of the selection of the RS type warp space as
our Standard Model throat in
the early universe. Second, we find several new dark matter
candidates. For example, we propose a novel type of dark matter --
the hidden dark matter -- which is generated during the tunneling
heating.
The hidden dark matter has many unusual properties compared to the
usual dark matter candidates.

Throughout the paper, we shall see the importance and richness of the
warped space and multi-throat compactification appearing in many places,
making them an even more attractive cosmological setup, in addition
to the various aspects already mentioned at the beginning of this
introduction.

To simplify the analysis, we ignore closed string modes other than the
gravity sector.
The lightest modes in the other relevant 
closed string sectors (such as moduli)
may either be more massive and stabilized throughout the whole
process; or become part of massive particle spectrum with similar
properties,
and their inclusion will not make a qualitative difference on the basic
issues.
Similar situations apply to the gravitino and its KK modes, 
so it is reasonable to expect that the
gravitino density will not pose a serious problem cosmologically in
this setup.
In general, the qualitative properties of a simplified warp metric is 
very close to that of a more realistic metric, such as that of the
Klebanov-Strassler (KS) throat \cite{Klebanov:2000hb}, whose KK properties 
have been studied \cite{Krasnitz:2000ir,Firouzjahi:2005qs}. 
The simplified metric of $S^{5}$ allows a simple analytic treatment so much 
of the physics is more transparent. In instances when the qualitative 
properties of the metric of the warped deformed conifold differ from that 
of $S^{5}$, we shall point that out explicitly.  

\subsection{Summary}

To make reading easier, here we summarize by giving a brief history
before the big bang nucleosynthesis according to the brane inflation
heating in the multi-throat scenario. For simplicity, here we only
keep the most important factors such as the warp factors $h_{A,S}$,
the length scales of the throats $R_{A,S}$, the string scale $m_s$,
the string couple $g_s$ and
the Planck mass $\mpl$. The subscript $A$ stands for the A-throat where 
brane annihilation towards the end of inflation takes place, while 
the S-throat is where the SM branes sit.

Immediately after the inflation at $t_0 \approx g_s^{1/2} h_A^{-2}
\mpl/m_s^2$, all branes and antibranes annihilate
in the A-throat and decay into heavy closed strings.
The universe enters into a period of matter-dominated
phase. These
closed strings subsequently decay into gravitons and their KK
modes. The branching ratio for graviton production is very small
because the KK modes are exponentially peaked at the infrared tip of the
A-throat, while the graviton, being the zero mode, is not
localized. In fact, our
estimate of the graviton production during this epoch 
is much smaller than that in 
Ref.~\cite{Kofman:2005yz} due to the bulk size suppression.

After a very short period $\Delta t\sim g_s^{-2} h_A^{-1} m_s^{-1}$,
the decay ends and then quickly thermalize into a tower of
relativistic KK particles at the bottom of the throat. 
The temperature is below the red-shifted string scale $h_A m_s$. 
The universe becomes radiation-dominated. 

As the universe expands, the temperature drops and the tower of KK
particles one by one become non-relativistic and then quickly fall out of
equilibrium. Because of warped space enhanced KK self-interaction, the
abundance of these frozen-out species are suppressed by the
warp factor and bulk size.

This lasts until $t \sim h_A^{-2} R_A^2 \mpl$ when the temperature drops 
below the lowest KK mode, $T \sim h_A R_A^{-1}$. The universe goes back 
to be matter-dominated.

These lowest KK modes, which stable against decay to gravitons, 
enter a long period 
of tunneling process to
other throats, in some cases going through the bulk resonance.
During this period, the KK particles are decoupled from each other
because of
the red-shifting of the spatial expansion. The graviton production is
negligible comparing to that in the early KK thermal history. Since
the graviton
density goes as $a^{-4}$ while the KK modes go like $a^{-3}$ ($a$ is
the scale factor), the graviton 
density rapidly becomes even more negligible.

Tunneling between two throats requires the energy level matching, and
the chance of matching becomes more frequent for longer throats, since longer
throats have denser spectra. Also because of this, the KK particles
tunneled into 
these throats quickly decay into lower KK levels and no longer tunnel
back. Due to these reasons, long throats are preferentially
heated during the tunneling. It is reasonable to assume that the S-throat, at TeV scale, 
is most warped throat or among the most warped. So this process heats the 
S-throat, as well as generating possible new dark matter candidates in
other throats. This long matter-dominated tunneling phase further
suppresses the final KK relics abundance in the A-throat and the
graviton abundance by delaying the Standard Model radiation-dominated phase.

By $t \sim \Gamma_{tun}^{-1}$, most energy has tunneled out of the
A-throat. Here the tunneling rate $\Gamma_{tun}$ can be of order
$h_A^9 R_A^{-1}$ or $h_A^{17} R_A^{-1}$ depending on 
whether the bulk resonance happens or not. For example, in
Ref.~\cite{Kachru:2003sx,Firouzjahi:2005dh,Shandera:2006ax},
matching brane inflation to
the CMBR data, the warp factor $h_{A}$ may range roughly from $1/4$ to
$10^{-3}$. This will yield a 
reasonable tunneling time for BBN.
In this case, the most heated throat turns out to be
the S-throat. The KK particles in this throat thermalize to a much
lower temperature and bring the universe back to be radiation-dominated.
At this moment, the temperature in the S-throat can be above or below
the local red-shifted string scale depending on the details.

A similar story then 
applies to the KK tower in the S-throat. The difference
is that, after the temperature drops below
the lowest KK mode at $t\sim h_S^{-2} R_S^2 \mpl$, the universe
remains radiation-dominated. This is because all the KK modes in
the S-throat finally release most of their entropy to the Standard Model 
particles on the branes at the bottom of the mass spectrum. 
The KK relics is similarly suppressed as those in the A-throat by the
bulk size and (much smaller) warp factors. This connects to the BBN.

The KK modes in the bulk will tunnel to the throats, mostly because, 
without warping, they are very massive, and their wavefunctions spread out 
throughout the bulk. So we do not expect any KK relics in the bulk for
the resonance case. We can also easily replace the S-throat with a bulk
SM brane and the similar story will apply after KK tunnels out of the
A-throat.

Our calculations suggest that KK modes may supply three types of new
dark matter candidates.

First, 
since the wave functions of KK modes with specific angular momentum in
the S-throat 
have no overlap (or very suppressed overlap) with SM branes sitting at the
bottom of the throat, they do not decay into SM particles. As a
result, they can be  
stable enough to become dark matter. 
Note that these are gravitational KK modes and that the Standard Model 
particles do not have KK excitations. So they are like the warped KK modes
\cite{Dimopoulos:2001ui,Dimopoulos:2001qd}, but  
different from the KK dark matter suggested in Ref.
\cite{Dienes:1998vg,Cheng:2002ej,Servant:2002aq}. 
They interact with the SM particles via the exchange of (unstable)
s-wave KK modes. 
In this sense, they are not too different from other dark matter
candidates, such as KK  
excitations of standard model particles or lightest supersymmetric
particles, though they  
can have very different production and scattering cross-sections. 

Second, KK modes in another long throat away from the SM branes
can also be  
dark matter candidates. If there are branes in such a throat, the KK
energy can be
converted to open string modes, much like what happens in the S-throat. 
Since these dark matters interact with SM particles only via
gravitons, 
they will be hidden dark matter. They can be detected via
gravitational lensing and rotation curves, but not
in collider experiments or underground dark matter searches.
These dark mater can have novel implications on the dark matter
coincidence problem.

Third, KK modes with specific angular momenta in the A-throat can also be
quite stable. Their tunneling rate is highly suppressed compared to the
s-wave KK modes that are responsible for tunneling in heating. This dark
matter is formed as thermal relics much like the first type of the KK
dark matter. But since they are decoupled from the SM particles except
for
graviton mediations, they also belong to hidden dark matter. Their  
masses are estimated to be very high, so their eventual tunneling to
the S-throat
and decay/interaction can produce very energetic events. Since these high 
energy events originate within the range of our galaxy but from
outside the SM branes, they can generate
cosmic rays that violate the GZK bound.

\medskip

This paper is organized as follows.

In Sec.~\ref{SecKKGKP}, we study the wave functions of the KK modes in
GKP type of string compactification. We also study the tunneling
properties of the KK particles between different throats. We estimate
the tunneling rates in different cases where the bulk resonance is
present or absent. 
 
In Sec.~\ref{SecKKinter}, we estimate the cross
sections of various possible interactions in the same setup. In both
Sec.~\ref{SecKKGKP} and \ref{SecKKinter}, some quantitative differences
between the GKP type and 5-d RS type setup are pointed out.

In Sec.~\ref{SecKKprod}, we outline the end process of the brane
inflation -- the
brane-antibrane annihilation -- and its various products. This leads
to the initial condition of our main study of the KK thermal history.

In Sec.~\ref{SecSingle}, \ref{SecDouble}, \ref{SecBulk} and
\ref{SecMulti}, we study
the thermal history of the KK particles in single throat, double
throat, bulk and
multi-throat cases, respectively. Estimates on the KK relic and
graviton density are made.
Interesting phenomena including the throat
deformation in the Hubble background, the
selection of long throats, the warped KK dark matter and the hidden
dark matter are discussed. 

In Sec.~\ref{SecHidden}, we study more
properties of the hidden dark matter, focusing on the dark matter
coincidence problem and possible implications on cosmic rays.

Sec.~\ref{SecDis} contains conclusions and discussions on future prospects.

\section{KK modes in GKP compactification}
\label{SecKKGKP}

In this section, we study the wavefunction and dynamics of KK modes
in throats. Previous studies on brane inflation heating have been
focused on the geometry of the orbifolded RS-type
\cite{Dimopoulos:2001qd,Barnaby:2004gg}. In this paper, we are
mostly interested in the generalized GKP-type
compactification
\cite{Giddings:2001yu,Kachru:2003aw,Verlinde:1999fy,Klebanov:2000hb},
namely, multiple throats with IR cutoffs whose UV sides are connected to a
six-dimensional bulk. We will first study the wavefunction and spectrum of a 
single throat case, then the tunneling of KK modes between different
throats.

\subsection{KK modes in a throat}
\label{SecKK}

In the GKP compactification, a KS throat is induced by three-form RR
and NSNS fluxes around a conifold singularity.
Except for the deformation around the IR cutoff of the throat, the
gravitational and RR background generated by fluxes are similar 
to that by D3-branes with the same D3-charge. 
Therefore the analyses of KK modes dynamics
is similar to that in the absorption cross section of
black holes which
have been studied extensively in the past
\cite{Starobinsky:1974,Gibbons:1975kk,Page:1976df,Unruh:1976fm,Dhar:1996vu,Das:1996we,Klebanov:1997kc}. 
The difference here is that
we will need to impose a different boundary condition, since we
prepare our initial state inside a throat.

We take the metric of the throat to be
\bea
ds^2 = A^{-1/2} (-dt^2 + d \bx^2 + h_{\mu\nu} dx^\mu dx^\nu) 
+ A^{1/2} (dr^2 + r^2 d\Omega_5^2 )~,
\eea
where we consider a simplified warp metric
\bea
A(r) = 1+ \frac{R^4}{r^4} ~.
\eea
The $r$ is the radial coordinate in six extra dimensions, and $R$
is the characteristic length scale of the warped space. The 4D KK
modes arise as metric perturbations $h_{\mu\nu}$, whose s-wave we
decompose as
\bea
h_{\mu\nu} = \tpsi_{\mu\nu}(x) \phi(r) ~.
\label{hmunu}
\eea
The leading order equation of motion for the s-wave 
KK modes is a Laplacian
equation in the background geometry,
\bea
\frac{1}{\rho^5} \frac{d}{d \rho} \left( \rho^5
\frac{d \phi}{d \rho} \right)
+ \left( 1+ \frac{(mR)^4}{\rho^4} \right) \phi = 0 ~,
\eea
where $m^2 = -p^2$ is the mass of the KK mode with $\tpsi_{\mu\nu}
\propto e^{ipx}$, and we have defined
$\rho \equiv mr$. 

To solve this differential equation, we redefine
\bea
\phi \equiv \rho^{-5/2} \tphi
\label{phidef}
\eea
and get
\bea
\left[ \frac{d^2}{d\rho^2} - \frac{15}{4\rho^2} 
+ 1+ \frac{(mR)^4}{\rho^4} \right] \tphi = 0 ~.
\label{tphieom}
\eea
This differential equation is of Schroedinger
type with zero energy, and the effect of the curved
background
geometry on KK modes is a potential barrier. 
We will study the inner and outer
regions of the barrier separately and match them.

The outer region is $\rho \gg mR$, where Eq.~(\ref{tphieom}) is
approximately 
\bea
\left[ \frac{d^2}{d\rho^2} - \frac{15}{4\rho^2} 
+ 1 \right] \tphi = 0 ~.
\label{tphieom1}
\eea
The solution is given by the Bessel functions
\bea
\tphi(\rho) = \rho^{1/2} \left( J_2(\rho) + i N_2(\rho) \right).
\label{tphiouter}
\eea
Here is the place where we have imposed the boundary condition,
that is, we only consider outgoing wave in the large $\rho$ region,
because
this is the KK mode leaking out of the throat. We can expand the
solution (\ref{tphiouter}) in two different limits. For $\rho \gg 1$, 
\bea
\tphi = \sqrt{\frac{2}{\pi}}~ e^{i(\rho - \frac{5\pi}{4})} 
~.
\eea
This describes an outgoing spherical wave. 
For $mR \ll \rho \ll 1$, 
\bea
\tphi = -i \frac{4}{\pi \rho^{3/2}} + \cdots 
+\frac{1}{8} \rho^{5/2} + \cdots ~.
\label{tphiouterB}
\eea
The first dots indicates the higher order imaginary terms
that are related to the $iN_2(\rho)$ term in
Eq.~(\ref{tphiouter}). The
leading real term  in Eq.~(\ref{tphiouterB}) comes from the
$J_2(\rho)$ term in Eq.~(\ref{tphiouter}).

The inner region is $\rho \ll mR$, where we can approximate
Eq.~(\ref{tphieom}) as
\bea
\left[ \frac{d^2}{d\rho^2} - \frac{15}{4\rho^2} +
\frac{(mR)^4}{\rho^4} \right] \tphi = 0 ~.
\eea
Changing variables, one can also bring this equation to Bessel's
differential equation and get
\bea
\tphi(\rho) = \frac{\rho^{1/2}}{mR} 
\left[ A J_2 \left( \frac{m^2 R^2}{\rho} \right)
+ B N_2 \left( \frac{m^2 R^2}{\rho} \right) \right] ~.
\eea
We can again expand this solution in two different limits.
For $m^2R^2 \ll \rho \ll mR$, 
\bea
\tphi = A \frac{m^3 R^3}{8 \rho^{3/2}} + 
\cO (\frac{m^7R^7}{\rho^{7/2}}) - 
B \frac{4\rho^{5/2}}{\pi m^5 R^5} + 
\cO ( \frac{\rho^{1/2}}{mR} ) ~.
\label{tphiinnerA}
\eea
Matching (\ref{tphiinnerA}) and (\ref{tphiouterB}) through $\rho \sim
mR$, we can determine the leading orders of the coefficients $A$ and
$B$ in Eq.~(\ref{tphiinnerA}) as
\bea
A &=& - \frac{32i}{\pi} m^{-3} R^{-3} ~, \nonumber \\
B &=& - \frac{\pi}{32} m^5 R^5  ~.
\eea
Note that there is also an imaginary part in $B$. 
We ignore this contribution
since our eventual goal is to calculate the leading tunneling
probability, in which such an imaginary term only leads to higher
order corrections. For $\rho \ll m^2 R^2$, 
\bea
\tphi = \sqrt{\frac{1}{2\pi}} \frac{\rho}{m^2R^2}
\left[ (A-iB) e^{i(\frac{m^2R^2}{\rho} - \frac{5\pi}{4})}
+ (A+iB) e^{-i(\frac{m^2R^2}{\rho} - \frac{5\pi}{4})} \right] ~.
\eea
This describes the outgoing and reflected incoming waves inside the
effective potential barrier. To the leading order, they have the
same amplitude. The small difference caused by the coefficient $B$
leads to the tunneling probability
\bea
P = 1- \left| \frac{A-iB}{A+iB} \right|^2 
= \frac{\pi^2}{2^8} m^8 R^8 ~.
\label{tunnelP}
\eea

We can perform a simple check on this tunneling probability by
comparing fluxes of the outgoing waves between the inner and outer
regions.
To do this, we list the leading behavior of the wave function $\phi$,
defined in (\ref{phidef}), up to an overall
normalization,
\bea
\rho \ll m^2R^2 &:& ~~~~~ 
\phi = -i \frac{32\sqrt{2}}{\pi^{3/2}} m^{-5} R^{-5} \rho^{-3/2} 
\cos \left( \frac{m^2R^2}{\rho} - \frac{5\pi}{4} \right)~, 
\label{wavetunnel1} \\
m^2R^2 \ll \rho \ll 1 &:& ~~~~~
\phi = - \frac{4i}{\pi} \rho^{-4} ~, 
\label{wavetunnel2} \\
\rho \gg 1 &:& ~~~~~
\phi = \sqrt{\frac{2}{\pi}} \rho^{-5/2} e^{i(\rho - \frac{5\pi}{4})}
~.
\label{wavetunnel3}
\eea
For $\rho \ll m^2R^2$ we have equal amount of outgoing and incoming
wave, the flux of the outgoing component can be calculated by
multiplying
the flux density ${\bf j} = \frac{1}{2i} (\phi_{\rm out}^* 
{\bf \nabla} \phi_{\rm out} -
c.c.)$ with the area of warped unit five-sphere, $\pi^3 A(r)^{5/4} r^5$,
and the warped unit longitudinal volume, $A(r)^{-1}$. Up to an overall
normalization, we get 
\bea
{\rm Flux} = \frac{\pi^3}{2i} \left( \phi^* r^5
\frac{\partial}{\partial r} \phi - c.c.\right) 
= 512~ m^{-12} R^{-8} ~.
\label{fluxt1}
\eea
The region $m^2R^2 \ll \rho \ll 1$ is the extension of the effective
potential, and the KK particles are tunneling. For $\rho \gg 1$, we
only have the tunneled outgoing KK particles. The corresponding flux is
\bea
{\rm Flux} = \frac{2\pi^2}{m^4} ~.
\label{fluxt2}
\eea
Comparing Eq.~(\ref{fluxt1}) and (\ref{fluxt2}), we get the tunneling
probability (\ref{tunnelP}). Here we have only considered the s-wave,
the tunneling probability of the higher partial waves will be
suppressed by factors of $mR$.

The mass quantization of the KK modes comes from the IR cutoff of the
throat at $\rho_0 = mr_0$.  Such a boundary condition will translate
into a constraint on the wavefunction (\ref{wavetunnel1}) or its first
derivative at $\rho_0$. The periodicity of the trigonometric function
then leads to $\Delta (m^2R^2/\rho_0) = \pi$. So the KK mass is quantized
in unit of 
\bea
\Delta m = \pi h_0 R^{-1} ~,
\eea
where 
\bea
h_0 = r_0/R
\label{warph0}
\eea
is the IR warp factor. The more realistic KS throat
has a more complicated smooth shape at the IR end, but we expect the
qualitative behavior of the mass quantization is
similar \cite{Firouzjahi:2005qs}.

\subsection{Tunneling between two throats}
\label{SecBetween}

KK modes in the bulk can also be absorbed by a throat. Such a process
is well analyzed previously when people study the black hole
absorption cross
section \cite{Starobinsky:1974,Gibbons:1975kk,Page:1976df,Unruh:1976fm,Dhar:1996vu,Das:1996we,Klebanov:1997kc}.
Here the boundary condition is to require
only incoming wave in the inner region. 
We list the leading behavior of the s-wave
in three different regions,
\bea
\rho \gg 1 &:& ~~~~~ 
\phi = -i \frac{32\sqrt{2}}{\pi^{3/2}}
m^{-5} R^{-5} \rho^{-5/2} \cos\left( \rho - \frac{5\pi}{4} \right) ~,
\label{waveabsorb1} \\
m^2R^2 \ll \rho \ll 1 &:& ~~~~~ 
\phi = -\frac{4i}{\pi} m^{-5} R^{-5} ~, 
\label{waveabsorb2} \\
\rho \ll m^2R^2 &:& ~~~~~ 
\phi = \sqrt{\frac{2}{\pi}} m^{-2}R^{-2}
\rho^{-3/2} e^{i ( \frac{m^2R^2}{\rho} - \frac{5\pi}{4} )}
~.
\label{waveabsorb3}
\eea
The tunneling probability turns out to be the same as
Eq.~(\ref{tunnelP}). This leads to the absorption cross section for
the s-wave \cite{Das:1996we,Klebanov:1997kc},
\bea
\sigma = \frac{\pi^4}{8} m^3 R^8 ~.
\label{absorbCS}
\eea

Now we imagine two throats -- A and X. The A-throat is where brane and
anti-brane annihilate and KK modes are generated. These KK modes can
leak through the bulk and tunnel into another throat -- X-throat,
which is originally empty. We assume that the A and X-throat are
separated by a distance $D$, and the bulk size is $L$. Let
$h_{A}=r_{A}/R_{A}$ and $h_{X}=r_{X}/R_{X}$ be the respective IR warp
factors. The tunneling between the two throats requires the matching of
the mass levels in two throats, $n_A h_A R_A^{-1} = n_X h_X
R_X^{-1}$, within the energy widths. Here $n_A$ and $n_X$ are the KK
levels in the A and X-throat, respectively.

We will consider two interesting cases. In the first case, consider
$L \gtrsim D \gtrsim m^{-1}$.
The size of the bulk are large enough,
so that the KK modes with mass $m$ may exist in the bulk. In this
case, we assume
such a mass quantization condition in the bulk is satisfied for $m$.
The KK modes will go in two
steps: first they tunnel out of the A-throat but remain in the bulk,
propagating and bouncing around (in the extra dimensions), then
gradually scatter and tunnel
into the X-throat. For the first step, because of the tunneling
probability (\ref{tunnelP}) and the fact that the KK mode in A-throat
has a bouncing period $2h_A^{-1} R_A$, the tunneling rate is
\bea
\Gamma_{A\rightarrow bulk} \approx \frac{\pi^2}{16^2} n_{A}^8 h_A^9
R_A^{-1} ~.
\label{rateAb}
\eea
For the second step, the absorption probability is given by the cross
section
\bea
P_{bulk\rightarrow X} \approx \frac{\sigma}{L^5} ~,
\eea
and the bouncing period of KK modes in the extra dimensions is $L$. So
the absorption rate is
\bea
\Gamma_{bulk\rightarrow X} \approx \frac{\sigma}{L^6} \approx
\frac{\pi^4}{8} n_{A}^3 h_A^3 \frac{R_X^8}{R_A^3 L^6} ~.
\label{ratebX}
\eea
The tunneling time for the KK particles to first go from the A-throat
to the bulk and then from the bulk to the X-throat is the summation of
these two tunneling time.
So the final rate is the smaller of these two rates (\ref{rateAb}) and
(\ref{ratebX}),
\bea
\Gamma_{A\rightarrow X} \approx {\rm Min} \left( \Gamma_{A\rightarrow
bulk}, \Gamma_{bulk \rightarrow X} \right) ~.
\label{trate1}
\eea
Because in (\ref{ratebX}), $L\gtrsim n_A^{-1} h_A^{-1} R_A$, we have 
$\Gamma_{bulk\rightarrow X} \lesssim
n_{A}^9 h_A^9 R_X^8/R_A^9$. 
So the tunneling rate for $n_A \sim 1$, $R_X \sim R_A$ is roughly 
\bea
\Gamma_{A\rightarrow X} \lesssim
h_A^9 R_A^{-1} ~.
\label{trateAXbulk}
\eea
The above analyses assumes that the energy of the KK modes in the
A-throat satisfies the mass quantization condition of the bulk, namely
the bulk KK mode also satisfies the mass matching condition as that
for the two throats. This
condition is the resonance effect that we will discuss in the next
subsection. 

The second case is that such a bulk mass quantization condition is not
satisfied. This also includes the case where the bulk size is too
small, $m< L^{-1}$. Assuming the A and X-throats are still well
separated, $D\gg R_A,R_X$, the KK particle has to tunnel two barriers
at one time, since they cannot propagate in
the bulk. Approximately, the tunneling
probability will be the product of the probability
(\ref{tunnelP}) for tunneling out and that for the absorption, and the
rate for $n_A \sim 1$ will be
\bea
\Gamma_{A\rightarrow X} \sim h_A^{17} R_A^{-1} ~.
\label{trateAXdir}
\eea
Of course one can always increase this tunneling rate by arranging
special configuration of throats, for example bringing them closer
and making the potential overlap. We will focus on the generic case.

\subsection{Resonance effect in tunneling}
\label{SecRes}

There is an equivalent way of looking at the same problem involving
the tunneling resonance effect in the presence of the bulk. Consider
the same setup in the
previous subsection.
There are two barriers that the particle must tunnel through:
the region between the A-throat and the bulk, $r_1 < r < r_2$, and the
region between the bulk and the X-throat, $r_3 < r < r_4$. The KK
modes have no barrier in the bulk, at $r_2 \le r \le r_3$.

The tunneling rate, or transmission coefficient, is straightforward to
obtain in the WKB approximation. 
Beginning with the first barrier, the A throat to the bulk, the
relation between the coefficients of the incoming waves and the
outgoing waves is given by (see Merzbacher \cite{Merzbacher})
\bea
\label{matrix1}
\frac{1}{2}\left( \begin{array}{c}
 \Theta_A  +\frac{1}{ \Theta_A}       \quad i\,(\Theta_A  -\frac{1}{ \Theta_A} )\\
-i\,(\Theta_A  -\frac{1}{ \Theta_A})  \quad  \Theta_A  +\frac{1}{ \Theta_A}
\end{array}  
\right) ~,
\eea
\bea
\Theta_A &=& 2\, \exp\left (\,\int_{\tau_1 }^{\tau_2} d\tau
\sqrt{V_{eff}(\tau)}\, \right ) ~,
\label{ThetaA}
\eea
where $\tau_1$ and $\tau_2$ are the classical  turning point.
Note that the effective energy in the analogous Schr{\"o}dinger
equation is zero. The variable $\tau$ and the effective potential is
defined in Ref.~\cite{Firouzjahi:2005qs}~.
The transmission coefficient from the A throat to the bulk is given by
\ba
\label{tran1}
P_{A \rightarrow bulk} &=&4 \left( \Theta_A  +\frac{1}{ \Theta_A}
\right)^{-2} \nonumber\\
&\simeq& \frac{4}{\Theta_A^{2}} ~.
\ea
This WKB method amounts to 
solving the leading tunneling behavior of the 
differential equation (\ref{phidef}) and (\ref{tphieom}). Indeed,
evaluating (\ref{ThetaA}), one obtains \cite{Firouzjahi:2005qs}
\ba
P_{A \rightarrow bulk} \sim (n_{A}\, h_A)^8 ~,
\ea
%where $h_A$ is the warp factor in A throat and $n$ is the radial KK mode.
in agreement with the tunneling probability (\ref{tunnelP}). 
   
Next we consider the probability of tunneling from the A-throat to
the X-throat via the bulk.
Up to some pre-factors, the matrix relating the 
coefficients of the incoming wave from the A-throat to the outgoing
wave in the X-throat is given by
\ba
\label{matrix2}
\frac{1}{4}\left( \begin{array}{c}
 \Theta_A  +\Theta_A^{-1}       \quad i\,(\Theta_A  - \Theta_A^{-1}  )\\
-i\,(\Theta_A  -\Theta_A^{-1})  \quad  \Theta_A  +\Theta_A^{-1}
\end{array}  
\right)  
\left( \begin{array}{c}
e^{-i\, W} \quad 0\\
0 \quad e^{i\, W} 
\end{array}  
\right)  
\left( \begin{array}{c}
 \Theta_X  +\Theta_X^{-1}       \quad i\,(\Theta_X  -\Theta_X^{-1})\\
-i\,(\Theta_X  -\Theta_X^{-1})  \quad  \Theta_X  +\Theta_X^{-1}
\end{array}  
\right) ~, \nonumber\\
\ea
where $W$ is the integral over the bulk
\ba
W=\int_{\tau_2}^{\tau_3} \sqrt{-V_{eff}(\tau)}\, d\tau ~.
\ea
For simplicity we assume $R_A =R_X$.
Even though the throats have different warp factors,
the WKB integral over the barrier between the bulk and the X-throat ends up to be the same 
as that from the A-throat to the bulk, $\Theta_X=\Theta_A$, since 
the KK quantization mode $n_{A}$ and $n_{X}$ obey $ n_A h_A R_A^{-1} =
n_X h_X R_X^{-1}$.

So the transmission coefficient from the A-throat to the X-throat via a bulk is given by
\ba
P_{A \rightarrow X} 
%& =& 4\, \left( ( \Theta_{A} \Theta_{S} + \frac{1}{\Theta_{A} \Theta_{S}})^{2} \cos^{2} L + (\frac{\Theta_{A}}{\Theta_{S}} +\frac{\Theta_{S}}{\Theta_{A}})^{2} \sin^{2} L\right)^{-1} \nonumber\\
=4\, \left( ( \Theta_{A}^2 + \frac{1}{\Theta_{A} ^2})^{2} 
\cos^{2} W + 4\, \sin^{2} W\right)^{-1} ~.
\ea
In the absence of the bulk, $W=0$ and $T (A \rightarrow S)$ is very small,
\ba
T_{A \rightarrow X} \sim \Theta_A ^{-4} \sim (n_{A}h_{A})^{16} ~.
\ea
This implies a tunneling rate
\ba
\Gamma_{A \rightarrow X} \sim  n_{A}^{16}h_{A}^{17}R_{A}^{-1} ~.
\label{tunnelWKB}
\ea

However, for KK modes satisfying
\bea
W=(n_{W}+1/2) \pi
\label{resonancecond}
\eea
so $\cos W =0$,
the transmission coefficient approaches unity 
\ba
T_{A \rightarrow X}  \sim 1 ~.
\ea
This is the well-known resonance effect. 
Notice that the location of the transmission peaks is determined by the same quantum 
condition for the bound states in the bulk.  For very massive KK
modes or large bulk, one finds
that the resonances easily occur.
This means that tunneling from the A throat to another throat passing
through the bulk may not be suppressed at all if the state is just at
the resonance point. However, in any realistic 
model, the initial KK wave packet
will have some widths which we have to average over. 
Suppose $h_{X} \ll h_{A}$. We expect a
$n_{A}$th KK mode will overlap with a small set of $n_{X}$th KK modes
so that the resonance effect will come into play. Or
if some KK modes in the A-throat are in thermal equilibrium, then it
is guaranteed that there are modes with the right energies to take
advantage of this resonance effect. In the following we estimate the
tunneling rate for these initial KK wave packets.

For large $\Theta_{A}$, so that the penetration through the barriers
is strongly suppressed, the transmission coefficient $T$ has sharp
narrow resonance peaks at these energies, with resonance width $\Delta
\Gamma$.
The distance between neighboring resonances is roughly 
\bea
D \simeq \frac{\pi}{\partial W/\partial E} ~.
\eea
Let us consider a wave packet that is localized in the A-throat at
$t=0$. To study the behavior of the transmitted wave packet near a
resonance, we take the initial wave packet to have a mean energy
$E_{0}$ corresponding to a resonance. The packet has a width $\Delta
E$ much bigger than the resonance width but much smaller than the
distance between neighboring resonances, i.e., 
\bea
D \gg \Delta E \gg \Delta \Gamma ~.
\eea
Around the resonance, 
\bea
\cos W \simeq \mp \left(\frac{\partial W}{\partial E}
\right)_{E=E_{0}} (E - E_{0})
\eea
and $\sin W \simeq 1$, so \cite{Merzbacher}
\bea
\sqrt{T} \simeq \mp \frac{{\Delta \Gamma}/2}{E - E_{0} + i {\Delta \Gamma}/2} ~,
\eea
where 
\bea
1/\Delta \Gamma = \frac{\Theta_{A}^{2}}{4} \left(\frac{\partial W}{\partial
E} \right)_{E=E_{0}} ~.
\eea
For $\Delta E \gg \Delta \Gamma$, the total transmission probability
for the incident wave packet is roughly
$\Delta \Gamma/\Delta E$. One may interpret this result in the
following way: the wave packet reaches the 
bulk in classical time.
A fraction $\Delta \Gamma/\Delta E$ of the packet is transmitted to
the X-throat according to the usual exponential decay law with a mean
lifetime $1/\Delta \Gamma$. The rest of the wave packet is reflected.

For the small band centered around the resonance point, using
\ba
\frac{\partial W}{\partial E} \sim L \gtrsim h_A^{-1} R_A ~,
\ea
we get the tunneling rate
\bea
\Gamma_{A\rightarrow X} \lesssim n_A^9 h_A^9 R_A^{-1} ~,
\eea
which agrees with (\ref{trateAXbulk}). Note that this is effectively
one barrier effect. This is because the resonance
condition (\ref{resonancecond}) is just the KK quantization condition
in the bulk, so if the resonance happens, the wave packet can stay and
bounce around in the bulk.

For a smooth distribution (e.g., a portion of a thermal distribution),
the efficiency of transmission depends on $\Delta \Gamma$, $\Delta
\Gamma/\Delta E$ and the time it takes  to replenish the transmitted
part of the wave packet. Assuming the thermalization time is much
shorter than the tunneling time,
the rate of transmission is  
\bea
\Gamma _{A \rightarrow X} \simeq \frac{\Delta \Gamma}{D} \Delta \Gamma
%\sim \frac{16R_{A}}{\pi^{3}\Theta_A^{4}h_{A}} 
\lesssim  n_A^{17} h_{A}^{17} R_A^{-1}~,
\eea
which agrees with that in Eq.~(\ref{tunnelWKB}) and
(\ref{trateAXdir}). This can also be understood -- a smooth
distribution overall does not satisfy the resonance condition.
% If the wave packet is part of the thermal spectrum, the replenish time is roughly $\lambda \beta$, where $\lambda$ is the coupling and $\beta$ is the inverse temperature.

In summary, for an initial KK wave packet in the A-throat (e.g.~with
$n_A=1$), if the
bulk resonance happens, the particles can propagate in the bulk and
the tunneling rate is enhanced for these particles,
$\Gamma_{A\rightarrow X} \sim
h_A^9 R_A^{-1}$. Otherwise, without resonance or for a smooth
distribution, the rate is much smaller, $\Gamma_{A\rightarrow X}
\sim h_A^{17} R_A^{-1}$. Notice that in either case, it is much
smaller than $\Gamma \sim h^5 R^{-1}$ for the RS 5-dimensional
space-time case.

\section{KK interactions}
\label{SecKKinter}

In this section, we study various cross sections of interactions among
KK modes, gravitons and brane fields. Since we eventually only need an
order-of-magnitude estimate of the thermal history, we will neglect
the numerical factors for simplicity.

The interactions among the KK particles and gravitons are determined
by the 10-dim action
\bea
\m10^8 \int d^{10} x \sqrt{-g_{10}} R_{10}
\supset \m10^8 \int d^5\Omega~ dr~ A ~r^5 
\int d^4x \sqrt{-g_4} R_4 ~.
\label{10daction}
\eea

We will first express various couplings in terms of the 10-d Planck
mass $\m10$ and then convert it to 4-d Planck mass $\mpl$. From the
action (\ref{10daction}), we can see that each throat
contributes $\m10^8 R_i^6$ to the 4-d Planck-mass-squared, 
while the bulk
contributes $\m10^8 L^6$. Since the typical situation in the
multi-throat compactification is that the size of the bulk $L$ is
bigger than the sum of length scales of all throats, $\sum R_i$, we
approximate
\bea
\mpl \approx \m10^4 L^3 ~.
\eea

Expanding (\ref{10daction}) in terms of the 4-d metric fluctuations
$h_{\mu\nu}$, we get
\bea
\m10^8 \int d^5\Omega~ dr~ A~ r^5 
\int d^4x \left( \partial h \cdot \partial h + 
h \cdot \partial h \cdot \partial h + 
h \cdot h \cdot \partial h \cdot \partial h +\cdots \right) ~,
\label{expansion}
\eea
where we only write the relevant expression schematically by neglecting 
indices that should be properly contracted with $\eta_{\mu\nu}$ and ignoring 
various numerical factors. 
After decomposing $h_{\mu\nu} =\sum \tpsi_{n \mu\nu}(x) \phi_{n}({\bf r})$ 
as in (\ref{hmunu}) and integrating out the wave functions
(\ref{wavetunnel1})-(\ref{wavetunnel3}) in the extra dimensions, we can
get couplings for the 4-d KK modes $\tpsi_{\mu\nu}$.

To make the kinetic term for the KK modes canonical, we first
integrate the extra dimensions for the quadratic term. It turns out
that the dominant contribution comes from the tip region $r\sim r_0$,
\bea
\m10^8 m^{-13} R^{-7} h_0^{-1} \int d^4 x (\partial \tpsi)^2 ~.
\eea
where $h_{0}$ is the warp factor (\ref{warph0}).
For the zero-mode graviton, the wave function in the extra dimension
is a constant, for example, we will take it to be one. 
The quadratic term is dominated by the bulk
\bea
\m10^8 L^6 \int d^4 x (\partial \tpsi_0 )^2 ~.
\eea
Therefore the canonically normalized fields are
\bea
\psi &\equiv& \m10^4 ~m^{-13/2} R^{-7/2} h_0^{-1/2} \tpsi ~, 
\nonumber \\
\psi_0 &\equiv& \m10^4 ~L^3 \tpsi_0 ~.
\label{canonical}
\eea

We next look at the interactions. We consider the case where all
KK particles and gravitons have similar energy. 
This applies to the case of the thermal equilibrium that we will focus
on later.
So in terms of orders
of magnitude, the 3-point interaction $\lambda_3 \int d^4 x~ \psi
\cdot \partial \psi \cdot \partial \psi$ corresponds to a cross
section $\sigma \approx \lambda_3^2$, and the 4-point interaction
$\lambda_4 \int d^4 x~ \psi\cdot\psi\cdot\partial\psi\cdot
\partial\psi$ corresponds to a cross section $\sigma \approx
\lambda_4^2 E^2$, where $E$ is the energy of the particles. We first
look at the cubic interactions.

{\it Three KK modes}: This is the KK self-interactions. Integrating
along the throat in (\ref{expansion}), the dominant
contribution comes from the tip of the throat, where the KK modes wave
function are peaked. In terms of the canonically normalized fields, 
we have
\bea
\m10^{-4} R^{-3} h_0^{-1} 
\int d^4 x ~\psi \cdot \partial \psi \cdot \partial \psi ~.
\eea
This corresponds to an interaction cross section
\bea
\sigma \approx \left( \frac{L}{R} \right)^6
\frac{h_0^{-2}}{\mpl^2} ~.
\eea

{\it Two KK modes and one graviton}: This is the KK modes annihilation
process. The coupling of this interaction is still dominated by
the tip region $r \sim r_0$,
\bea
\m10^{-4} L^{-3} \int d^4 x ~
\psi \cdot \partial \psi \cdot \partial \psi ~,
\eea
where one of $\psi$ is $\psi_0$. This leads to an annihilation cross
section
\bea
\sigma \approx \frac{1}{\mpl^2} ~.
\label{csKKg}
\eea

{\it One KK mode and two gravitons}: Because of the KK momentum
conservation, namely
\bea
\int dr~ r^5 A~ \phi(r) = 0 ~,
\eea
such a cross section vanishes. This equation is in fact the
orthogonality condition between the KK and the zero mode, so 
this conclusion can be easily generalized to more arbitrary geometry or
multi-gravitons. Since any compactification 
has a zero mode graviton with constant wavefunction
\cite{Firouzjahi:2005qs}, $n$-point function involving one KK mode and
$n-1$ gravitons is proportional to the orthogonality condition between
the KK and graviton mode. Hence the KK mode cannot decay to all gravitons
directly.
This conclusion still holds when all other terms in the action
(\ref{10daction}) are considered. The arguments are given in Appendix
\ref{AppKKdecay}.
For higher KK mode, it can still decay to several lower KK
modes plus graviton with a rate $\ll m^3/\mpl^2$. 
But for the lowest KK mode, the only way it can
emit graviton is to annihilate or scatter 
another particle.

{\it Three gravitons}: The graviton self-interactions are dominated by
the bulk, their coupling and cross section turns out to be the same as
the KK annihilation case, $\sigma \approx 1/\mpl^2$.

If branes exist at the IR tip of the throat, the open string fields on
the brane will interact with KK particles and gravitons.
These interactions are given by
the following term in the action,
\bea
\int d^4 x d^5 \Omega dr~ A^{1/2} \sqrt{-g}
~\half~ G^{\mu\nu} \partial_\mu \tH \partial_\nu \tH 
~\delta(r-r_0) ~\delta^5(\Omega-\Omega_0) ~,
\label{actionH}
\eea
where $G^{\mu\nu} = A^{1/2} g^{\mu\nu}$, $r_0$ and $\Omega_0$ are the
radial and angular location of the D3-branes. The $\tH$ denotes the brane
fields, which for
example can be a Standard Model Higgs.
Expanding (\ref{actionH}) in terms of the
4-d fluctuations $h_{\mu\nu}$, one gets
\bea
\frac{R^2}{2 r_0^2} \int d^4 x \left( \eta^{\mu\nu} \partial_\mu \tH
\partial_\nu \tH + h^{\mu\nu} \partial_\mu \tH \partial_\nu \tH +
h\cdot h \cdot \partial \tH \cdot \partial \tH + \cdots 
\right)_{r=r_0} ~.
\label{actionHexp}
\eea
We denote the canonically normalized SM particle
as 
\bea
H \equiv \frac{R}{r_0} \tH ~.
\eea
The presence of the D3-branes breaks the conservation of the KK
momentum. So for example, a KK particle can hit the brane and break
into two SM particles.

{\it KK-SM interaction}: Such a three-point interaction is
determined by the second term in (\ref{actionHexp}),
\bea
M_{10}^{-4} R^{-3} h_0^{-1} \int d^4 x
\tpsi^{\mu\nu} \partial_\mu H \partial_\nu H ~.
\label{psiHH}
\eea
The cross section is 
\bea
\sigma \approx \left( \frac{L}{R} \right)^6 
\frac{h_0^{-2}}{\mpl^2} ~.
\eea
If the KK particle is replaced by a graviton, we get
\bea
\sigma \approx \frac{1}{\mpl^2} ~.
\label{sigmagHH}
\eea

\medskip
The quartic interactions can be analyzed in a similar way. In the
following, we summarize both the cubic and quartic interactions
according to the interaction type.
In all cases, the cubic interactions are the most important.

\begin{itemize}
\item {\it KK self-interactions}:
This includes $KK + KK \rightarrow KK$, $KK + KK \rightarrow KK + KK$,
$KK +KK \rightarrow KK + g$ and higher points interactions. 
The cross section of these three interactions are given by the three
terms in 
\bea
\sigma_{KK} \approx 
\left( \frac{L}{R} \right)^6 
\frac{h_0^{-2}}{\mpl^2}
\left( 1+ \frac{n_E^2}{N^2} + \frac{n_E^2}{N^2} \left( \frac{R}{L}
\right)^6 h_0^2 + \cdots \right) ~,
\label{csKK1}
\eea
respectively.
Here $n_E$ parameterizes
the particle energy scale, $E = n_E h_0 R^{-1}$.
For relativistic species, $n_E > n_{KK}$, and for non-relativistic
species, $n_E = n_{KK}$.
In the last two terms, the relations $R^4 \approx N g_s/m_s^4$ and
$\mpl^2 \approx m_s^8 L^6/g_s^2$ have been used. These two terms are
negligible to the first, so we have
\bea
\sigma_{KK} \approx 
\left( \frac{L}{R} \right)^6 
\frac{h_0^{-2}}{\mpl^2} ~.
\label{csKK}
\eea
Comparing to the
RS case, we have the same warp factor enhancement $h_0^{-2}$. In
addition, the cross section in the GKP case is further enhanced by the
bulk size $L^6/R^6$.

\item {\it Hard graviton production from KK annihilation}:
This includes $KK + KK \rightarrow g$, $KK +KK \rightarrow KK + g$,
$KK + KK \rightarrow g+g$ and higher points interactions. 
The cross section for the first reaction is
(\ref{csKKg}), and the third has an additional
suppression factor $E^2/\mpl^2$. The second process does not entirely
annihilate the KK modes and was included in the previous case as a KK
self-interaction. But this reaction emits a hard graviton and can be
interesting as we consider the graviton production. The cross section
is $\frac{n_E^2}{N^2} \frac{1}{\mpl^2}$. 
So the first reaction dominates,
\bea
\sigma_{Kg} \approx \frac{1}{\mpl^2} ~.
\label{csKg}
\eea

\item {\it Graviton interactions}: 
This includes $g+g \rightarrow g$, $g+g
\rightarrow g+g$, $g+g \to H + H$, 
$g+g \rightarrow KK + KK$ and higher points interactions. 
The largest cross
section is that of the first,
\bea
\sigma_{gg} \approx \frac{1}{\mpl^2} ~.
\label{csgg}
\eea
The latter three are suppressed by an additional factor of
$E^2/\mpl^2$. So as usual
the graviton interaction is Planck-mass suppressed.

\item {\it KK-SM interactions}:
This includes $H+H \rightarrow KK$, $H+H \rightarrow KK + KK$ and
higher points interactions. 
Their cross section is
\bea
\sigma_{HK} &\approx&
\left( \frac{L}{R} \right)^6 
\frac{h_0^{-2}}{\mpl^2}
\left( 1+ \frac{n_E^2}{N^2} + \cdots \right)
\nonumber \\
&\approx&
\left( \frac{L}{R} \right)^6 
\frac{h_0^{-2}}{\mpl^2}
\label{csHK}
\eea

\item {\it Hard graviton production from SM particles
annihilation}:
This includes $H+H \rightarrow g$, $H+H \rightarrow g+KK$, $H+H
\rightarrow g+g$ and higher points interactions. 
Their cross sections are
\bea
\sigma_{Hg} &\approx& \frac{1}{\mpl^2} \left( 1+ \frac{n_E^2}{N^2} +
\frac{E^2}{\mpl^2} +\cdots \right) 
\nonumber \\
&\approx& \frac{1}{\mpl^2} ~.
\label{csHg}
\eea

\end{itemize}

We note that the wavefunction (\ref{wavetunnel1})-(\ref{wavetunnel3})
used here is for s-wave. For higher partial wave, we will have a phase
shift in the wave function and an extra angular part. The qualitative
estimates we made in this section also apply to those KK modes with 
non-vanishing wavefunction at the bottom of the throat where the SM 
branes are sitting, such as those associated with $S^3$ 
in a realistic deformed warped conifold of the 
KS throat. There, 
the KK modes with $S^{2}$ angular momentum has vanishing 
wavefunctions at the origin, so they will interact 
with the SM particles via the exchange of the unstable s-wave KK modes.

\section{Production of KK modes from brane annihilation}
\label{SecKKprod}

To study the heating process in brane inflation, we first look at the
initial state from the brane-antibrane annihilation. 
We choose the time coordinate by choosing $\rho \approx
\mpl^2/t^2$. So at the initial brane annihilation $\rho \approx h_A^4
T_3$, we have $t_0 \approx g_s^{1/2} h_A^{-2} \mpl/m_s^2$. 
The string
dynamics in brane annihilation can be described by a boundary
conformal field
theory \cite{Sen:2002nu,Sen:2002in,Sen:2004nf}. Calculations in the
tree level \cite{Lambert:2003zr} and loop
level \cite{Chen:2003xq,Leblond:2005km}
string amplitudes
indicate that the dominant product is non-relativistic massive closed
strings, no matter whether there are any branes left over from such an
annihilation. These heavy closed strings will further decay to lighter
states, such as lighter closed and open strings, gravitons and KK
modes. The time
scale for such annihilation and decay processes is of order
red-shifted string scale, $\Delta t \sim g_s^{-2} h_A^{-1} m_s^{-1}$,
up to a constant factor depending on string levels and configurations
\cite{Iengo:2006gm}. The $\Delta t$ is much smaller than $t_0$, so
these decay happens right after $t_0$. 

As we will check later (in Eq.~(\ref{thermalcond})), 
it is easy to satisfy the condition 
that the spatial expansion
of the universe is slow enough for these products to thermalize. 
So both the decay and thermalization time is smaller than
$t_0$. Therefore the initial energy density of the thermalized gas is
approximately the same as the initial brane tension $h_A^4 T_3$.
For a gas of
relativistic particles with $g$ degrees of freedom, the thermalized
temperature will be $T_0 \approx (gg_s)^{-1/4} h_A m_s$. 
For $g>g_s^{-1}$ this is below
the red-shifted string scale in the throat. Hence these
thermalized particles should be
the graviton KK modes if $T_0 > h_A R_A^{-1}$, 
as well as the brane fields 
(zeroth open strings
modes) if there are branes left.

Gravitons are also present as a decay product. Since their cross
section is of order $\mpl^{-2}$, they are never in thermal equilibrium
as long as time is larger than the Planck scale. So it is already
important to estimate its initial abundance.

If the brane-antibrane annihilation are in the bulk,
the graviton and the bulk KK modes couplings to the heavy closed
strings are of the same order of magnitudes. Then the initial graviton
density will be a sizable fraction of the total energy density. Since during
the BBN, graviton density cannot exceed a few percent of the total
energy density of the universe, a more detailed analyses is required
to determine this fraction. The analyses in
Ref.~\cite{Chialva:2005zy} indicates
that this is problematic. (However see a solution at the end of
Sec.~\ref{SecBulk}.)

The situation is very different in the warped space. This is because,
unlike the zeroth mode graviton, KK modes are peaked
at the tip of the warped space. In the 5-dim RS setup, their coupling
to the heavy strings located at the IR end of the warped space will be
enhanced by
a factor of $h_A^{-1}$ relative to the graviton. Hence the graviton
production is suppressed by a factor of
$h_A^2$ \cite{Barnaby:2004gg,Kofman:2005yz,Chialva:2005zy}.  
The case that we consider here in the GKP setup is essentially the same 
except for the quantitative details. From the analyses in
Eqs.~(\ref{psiHH})-(\ref{sigmagHH}), we know that the suppression for the
graviton production is greater -- by a factor of
\bea
\left ( \frac{R}{L} \right)^6 h_A^2 ~. 
\label{GravSupp}
\eea
To get an order of magnitude feeling,
we note that, in fitting CMBR data, $R/L \sim 1/3$ to $1/100$  and 
$ h_{A} \sim 1/4$ to $1/1000$ are reasonable for inflation in A-throat
\cite{Kachru:2003sx,Firouzjahi:2005dh,Shandera:2006ax}.
If inflation takes place as the D3-brane is coming out of another
throat, as in
the IR DBI case \cite{Chen:2004gc,Chen:2005ad}, then the value of $h_{A}$ is 
more relaxed.

So in the presence of any reasonable warp factor, 
the dominant products after brane-antibrane
annihilation, string decay and particle 
thermalization is a tower of relativistic KK
modes with possible brane fields at the bottom of the mass
spectrum.\footnote{The initial KK can be absent if $T_0<h_A
R_A^{-1}$, so we only have relativistic brane fields. This simplified
case only
happens for very small $N_A < g$ in the single throat case.}
The thermal history of such a gas will be the main focus of our next few
sections. The study of the evolution of such states will eventually
lead us to conditions at the time of BBN, and hence provides a test
whether this is a viable heating.

\section{Single throat heating}
\label{SecSingle}

In this section, we study the time-dependent thermodynamics of KK
modes in single throat case. In this case, the D3-anti-D3-brane
annihilation, and therefore their eventual product KK particles, are
in the same throat as the Standard Model branes. Our goal is to
give an estimate of the relic KK mode and graviton production from the
thermal history. To keep it more general, in this and next section, we
will first obtain results in terms of general cross sections and tunneling rate, 
and then evaluate them using the results in GKP type compactification
in Sec.~\ref{SecKKGKP} \& \ref{SecKKinter}. It can also be evaluated
in the Randall-Sundrum setup.

Initially the Universe is dominated by a gas of relativistic KK modes
that are created from brane and string decay. We know that the
evolution of the energy density during such a period follows
\bea
\rho = \frac{3 \mpl^2}{4 t^2} ~.
\label{rhot}
\eea
If the KK particle self-interacting rate $\Gamma(t)$ is bigger than the
Hubble expansion rate $H(t)$, the KK particles are in thermal
equilibrium. We will check shortly that this is the case for the
period that we are interested in. The density of the thermalized
relativistic particles is determined by the temperature,
\bea
\rho \approx \frac{\pi^2}{30} g T^4 ~,
\label{rhoT}
\eea
where $g$ is the number of degrees of freedom of all relativistic
particles in
equilibrium. These particles include the KK and SM particles and their
possible super-partners. For
simplicity, we will take $g$ to be a constant of the same order of
magnitude as the SM particles $\sim 100$. Variation of $g$ may come if
the KK tower is very high at the early epoch, or in the absence of the
SM particles. We will encounter such cases in the next section.
The difference caused by the variation in $g$ is not essential in this
section.

The time dependence of temperature $T$ then follows from
Eq.~(\ref{rhot}) and (\ref{rhoT}),
\bea
T(t) \approx g^{-1/4} \left( \frac{\mpl}{t} \right)^{1/2} ~.
\label{Tt}
\eea
As the universe expands, the temperature decreases. Each KK mode
consecutively becomes non-relativistic as $T$ drops below its
mass. 

At $T_{nr} \approx m = n_{KK} h_0 R^{-1}$, the temperature crosses the
mass of the $n_{KK}$th KK
modes (including all possible degenerate states) and these KK
particles are becoming non-relativistic. From
(\ref{Tt}), the corresponding time is
\bea
t_{nr} \approx g^{-1/2} \mpl m^{-2} ~.
\label{tnr}
\eea
The condition that we can use the thermodynamics for this KK mode
until this moment is
\bea
\Gamma/H \approx \frac{n \sigma_{KK} v}{1/(2t_{nr})} \approx
g T^3 \sigma_{KK} t_{nr} \gtrsim 1 ~,
\label{thermalcond}
\eea
where $n$ is the number density of this KK mode and the velocity $v$
of the KK particle is approximated as $1$. This is easy to
satisfy. For example, for $\sigma_{KK}$ in (\ref{csKK}), 
\bea
\Gamma/H \approx g^{1/2} n_{KK} R^{-1} \mpl^{-1} \left( \frac{L}{R}
\right)^6 h_0^{-1} ~.
\eea
If $g^{1/2} \approx 10$, $n_{KK} \approx 1$, $m_s \approx 10^{10} {\rm
GeV}$, $N_A^{1/4} \approx 10$ ($R_A^4 \approx g_s N_A m_s^{-4}$), 
and $\mpl \approx 10^{18} {\rm GeV}$,
we need $\left( \frac{L}{R} \right)^6 h_0^{-1} > 10^8$. This can be
easily satisfied e.g.~by $L\gtrsim 10 R$ and $h_0 \lesssim 0.01$. 

The density of these non-relativistic KK particle will drop
exponentially with temperature
\bea
n_{nr} = g\left( \frac{m T}{2\pi} \right)^{3/2}
e^{-m/T} ~,
\label{nnonrel}
\eea
and their entropy is released to the other relativistic species.
The self-interacting rate of such a particle also drops accordingly
and this species quickly becomes decoupled. 
After $t_{nr}$, the decrease of the interaction rate $\Gamma =
n_{nr} \sigma_{KK} v$ is dominated by the exponential decrease of
$n_{nr}$. So when the KK particle is decoupled, $\Gamma \approx
H(t_{dec})$, we can estimate the leading order of the particle number
density as 
\bea
n^{KK}_{dec} \approx H(t_{dec}) \sigma_{KK}^{-1} v^{-1}(t_{dec})
\approx H(t_{nr}) \sigma_{KK}^{-1} 
\approx g^{1/2} \mpl^{-1} m^2 \sigma_{KK}^{-1} ~.
\label{nKKdec}
\eea
Here the changes in $H$ and $v$ from $t_{nr}$ to $t_{dec}$ are
neglected for simplicity.\footnote{Using (\ref{nnonrel}), we can
estimate $T(t_{dec})
\approx m \left( \ln \frac{g m^3}{n_{dec}} \right)^{-1}$. So,
the velocity $v$ at $t_{dec}$ is $v \approx \sqrt{ T/m} \approx 
\left( \ln \frac{gm^3}{n_{dec}} \right)^{-1/2}$, which changes
logarithmically slower than $n_{KK}$, consistent with our
assumption. The same happens to $H$. An overall correction
is a factor of $\left( \ln \frac{g m^3}{n^{KK}_{dec}} \right)^{-3/2}$,
which is smaller than one and makes the $n^{KK}_{dec}$ in
(\ref{nKKdec}) smaller.} These changes will give further suppression
factor which is logarithmically weaker.

The whole tower of KK modes is gradually lowered in this fashion, 
until the temperature drops 
below the mass of the last KK mode $h_0 R^{-1}$. At this point, all KK
particles are non-relativistic and release most of their entropy to
the relativistic SM particles.

If any KK modes are stable because they carry some conserved quantity, 
they become relics after decoupling. It
is important to estimate their relative abundance to the other
relativistic SM particles. Since non-relativistic KK particles
red-shift differently from
relativistic SM particles, this relative abundance will be amplified
in later evolution. This amplification lasts until the transition
epoch from 
radiation domination to matter domination (RDMD), at which time
the density of these non-relativistic
relics cannot
be more than that of the
radiations, even if they account for all the dark matter.

We now estimate this relative ratio.
At $t_{dec}$, the ratio of the energy density of the
decoupled KK particle to the radiation energy density is 
\bea
\left( \frac{\rho^{KK}}{\rho^{tot}} \right)_{t_{dec}}
\approx \frac{n_{dec} m }{ g m^4}
\approx \frac{1}{g^{1/2} m \mpl  \sigma_{KK}} ~.
\eea
Summing over all possible stable KK modes and taking into account of
the red-shift, at the time of the
radiation-matter transition, we have
\bea
\Omega_{RDMD}^{KK} 
&\equiv&
\left( \frac{\rho^{KK}}{\rho^{tot}} \right)_{RDMD}
\nonumber \\
&=& \sum_{n_{KK}} 
\left( \frac{\rho^{KK}}{\rho^{tot}} \right)_{t_{dec}}
\frac{ a(t_{RDMD}) }{ a(t_{dec}) } \nonumber \\
&=& \sum_{n_{KK}}
\left( \frac{\rho^{KK}}{\rho^{tot}} \right)_{t_{dec}}
\left( \frac{\rho^{tot}(t_{dec})}{\rho_{RDMD}} \right)^{1/4}
\nonumber \\
&\approx& \sum_{n_{KK}}
g^{-1/4} \mpl^{-1} \rho_{RDMD}^{-1/4} \sigma_{KK}^{-1}~,
\label{KKratio}
\eea
where $\rho_{RDMD}$ is the energy density at the time of RDMD
transition.
Using the cross section $\sigma_{KK}$ in (\ref{csKK}), we
have\footnote{The cross section of the KK self-interaction can be
enhanced by the KK-SM particles interactions (\ref{csHK}). Especially
when we have many different SM particles. Such an enhancement can
further decrease the ratio.}
\bea
\Omega_{RDMD}^{KK} 
%\equiv
%\left( \frac{\rho^{KK}}{\rho^{tot}} \right)_{RDMD}
\approx \sum_{n_{KK}} g^{-1/4} \left( \frac{R}{L} \right)^6 
\frac{\mpl}{\rho_{RDMD}^{1/4}} h_0^2 ~.
\label{relics}
\eea
For the single throat case, this ratio can be made very small if $h_0$
is small.
Since $\rho_{RDMD}^{1/4} \approx {\rm 1~eV}$, $\mpl \approx 10^{18}
{\rm GeV}$, we need $\left( \frac{R}{L} \right)^6 h_0^2 \lesssim
10^{-27}$. For example, this can be satisfied if $L\approx 100 R$ and
$h_0 \approx 10^{-8}$. 
A few comments are in order here. First, we can also evaluate
(\ref{KKratio}) in the 5-d RS case. It is still suppressed by the warp
factor $h_0^2$ but without the bulk size suppression. 
So the condition becomes
more restrictive, $h_0 \lesssim 10^{-14}$.
Second, as we will see in the next section,
$\Omega_{RDMD}^{KK}$ will receive another suppression factor 
in the double throat
case due to an extra matter-dominated phase. Third, we note
that we have been interested in the 
relic density resulted from the thermal history. Whether these relics
can be stable enough is also important and case dependent
\cite{Kofman:2005yz,DeWolfe:2004qx}. In this paper, we show that even
the 
most restrictive case, namely the stable relics, is compatible with a
successful heating.

We now estimate the graviton production. They are produced through the
KK and SM particles annihilation. Once produced, they are always
decoupled and red-shifted as radiation. Therefore their density is
\bea
\rho^{grav}(t) \approx \int_{t_0}^{t} 
T(t_1)~ n^2(t_1)~ \sigma_{Kg} ~\frac{a^4(t_1)}{a^4(t)}~ dt_1 ~,
\label{gravprod}
\eea
where we estimate the graviton energy as $T$, which is the energy
scale of the reactions involved. The $n\approx g T^3$ is the number
density of the KK or SM particles. From (\ref{csKg})
and (\ref{csHg}) we know that $\sigma_{Hg} \approx \sigma_{Kg}$. 
The initial time $t_0$ is set by initial energy density released from
the brane annihilation, $H_0^2
\approx \frac{1}{4t_0^2} \approx \frac{T_3 h_0^4}{3\mpl^2}$, as in
Sec.~\ref{SecKKprod}.
So we get
\bea
\rho^{grav}(t) \approx g_s^{-1/4} g^{1/4} \mpl^3 m_s \sigma_{Kg} h_0
\frac{1}{t^2} ~.
\label{gravprod0}
\eea
In this integration, we can see that
the dominant contribution comes from the initial moment $t_0$ over a
period of order $t_0$. After that, $\rho^{grav}$ just red-shifts as
$1/t^2$. Therefore the faction of gravitons in total radiation is
\bea
\Omega_{BBN~or~RDMD}^{grav} \equiv
\left( \frac{\rho^{grav}}{\rho^{tot}} \right)_{BBN~or~RDMD}
\approx g_s^{-1/4} g^{1/4} \mpl m_s \sigma_{Kg} h_0 ~.
\label{gravratio}
\eea
Using the cross section (\ref{csKg}), we have
\bea
\Omega_{BBN~or~RDMD}^{grav} 
%\equiv
%\left( \frac{\rho^{grav}}{\rho^{tot}} \right)_{BBN~or~RDMD} 
\approx 
g_s^{-1/4} g^{1/4} \frac{m_s}{\mpl} h_0 ~,
\label{gravratio2}
\eea
which is naturally very small.\footnote{Higher KK particle can also
decay to several lower KK particles plus graviton with a rate $\ll
m^3/\mpl^2$. Within a time interval $t_{nr} \approx g^{-1/2} \mpl
m^{-2}$, the decayed fraction is $\ll g_s^{-1/4} g^{-1/2} N^{-1/4}
\frac{m_s}{\mpl} h$, much smaller than (\ref{gravratio2}).}
We note that the ratio
(\ref{gravratio}) is proportional
to the graviton production cross section $\sigma_{Kg}$, but is
independent of
the KK self-interactions cross section $\sigma_{KK}$. This is the 
difference between the thermal dynamics and simple reactions. That is,
KK and SM particles are in thermal equilibrium and their density is
determined by the temperature rather than the cross section, while the
gravitons are never in thermal equilibrium. The total graviton
abundance is the summation of (\ref{GravSupp}) and
(\ref{gravratio2}).

\section{Double throat heating}
\label{SecDouble}

In this section, we consider the heating process in double throat
case. In this case, the brane-anti-brane annihilation happens in the
A-throat and there is no brane left in the end. The Standard Model
lives on the branes in the S-throat. So successful heating requires
efficient energy transfer from KK modes in the A-throat to branes
in the S-throat. We study this process in three steps: the evolution
of KK modes in the A-throat, the tunneling between A and S-throat and
the evolution of KK modes in the S-throat.

\subsection{KK evolution in A-throat}
\label{SecKKinA}

The thermal history of the KK modes in the A-throat is very similar to
that in the single throat case. The main difference is that at the
bottom of the spectrum there is no longer SM particles. So even if the
KK tower starts not very high, the particle degrees of freedom will
not remain constant. We write the relation between $g$ and the
temperature $T$ in the following ansatz
\bea
g \approx n_{KK}^\gamma \approx \left( \frac{T}{h_A R_A^{-1}}
\right)^\gamma
\approx \tilde\alpha^{1+\frac{\gamma}{4}} 
\left( \frac{T}{h_A T_3^{1/4}} \right)^\gamma ~,
\label{gTA}
\eea
where the value of $\gamma$ depends on the degeneracy of the
states. For example, for 5-d RS case, $\gamma =1$; for $AdS_5 \times
S^5$, $\gamma \approx 5$; for $AdS_5 \times T^{11}$, $\gamma \approx
4$ to $5$.
Sometimes, as in the last step, we will also write $R$ in
terms of brane tension $T_3$, where $\talpha \approx
N^{\frac{\gamma}{\gamma+4}}$. 
Therefore the thermal density of the KK modes is
\bea
\rho(t) \approx \frac{T^{\gamma+4}}{(h_A R_A^{-1})^\gamma} ~,
\label{rhotTt}
\eea
and the time dependence of the temperature follows
\bea
T(t) \approx \left( h_A R_A^{-1} \right)^{ \frac{\gamma}{\gamma+4} }
\left( \frac{\mpl}{t} \right)^{\frac{2}{\gamma+4}} ~.
\label{TtA}
\eea
The KK modes at each level $n_{KK}$ become non-relativistic when the
temperature drops below $m \approx n_{KK} h_A R_A^{-1}$ at 
\bea
t_{Anr} &\approx& \left( h_A R_A^{-1} \right)^{\frac{\gamma}{2}} \mpl
m^{-\frac{\gamma+4}{2}} 
\approx n_{KK}^{-\frac{\gamma+4}{2}} R_A^2 \mpl h_A^{-2} \nonumber \\
&\approx& t_{Adec} ~.
\label{tAnr}
\eea
The decoupling time $t_{Adec}$ for these KK modes follows shortly
after $t_{Anr}$.
The main difference between this case and the previous single throat
case is that,
as the temperature drops below the
last KK mode in the A-throat, the universe enters a period of matter
dominated (MD) phase. The lowest KK mode stays in the A-throat waiting
to tunnel. The tunneling time $t_{tun}\approx \Gamma_{tun}^{-1}$ is
estimated in 
Sec.~\ref{SecKKGKP}. 

The higher partial waves of KK modes take
much longer time to tunnel, and if stable enough, can become
non-relativistic relics.
This relics density can be estimated at the time when they become
decoupled. The corresponding number density is
\bea
n_{Adec} \approx \frac{H(t_{Anr})}{\sigma_{KK}} \approx t_{Anr}^{-1}
\sigma_{KK}^{-1} ~. 
\eea
The relative density of these KK relics at decoupling is
\bea
\left( \frac{\rho^{KK}}{\rho^{tot}} \right)_{t_{Adec}} \approx
\frac{ n_{Adec} m}{ \mpl^2/t_{Adec}^2} ~.
\eea
This ratio will be amplified during any radiation domination epoch
that follows. As we will see, this will start from when the tunneling
is done ($\approx t_{tun}$) and lasts until the RDMD
transition in the S-throat. So at RDMD,
the relics abundance is
\bea
\Omega_{RDMD}^{KK} \equiv
\left( \frac{\rho^{KK}}{\rho^{tot}} \right)_{RDMD}
&\approx&
\sum_{n_{KK}}
\left( \frac{\rho^{KK}}{\rho^{tot}} \right)_{t_{Adec}}
\frac{\mpl^{1/2} t_{tun}^{-1/2}}{\rho_{RDMD}^{1/4}} 
\nonumber \\
&\approx&
\sum_{n_{KK}} n_{KK}^{-\gamma/4} \mpl^{-1} \rho_{RDMD}^{-1/4}
\sigma_{KK}^{-1} \left(\frac{t_{Anr}}{t_{tun}}\right)^{1/2} ~.
%\sum_{n_{KK}} n_{KK}^{3-\frac{n}{4}} \mpl \left( \frac{R_A}{L}
%\right)^6
%\frac{h_A^4}{\rho^{1/4}_{RDMD}}
%\left(\frac{t_{Anr}}{t_{tun}}\right)^{1/2} ~.
\label{relicsA}
\eea
The main difference between Eq.~(\ref{relicsA}) and (\ref{KKratio}) is
the last factor in Eq.~(\ref{relicsA}). This further
suppression is due to the matter dominant phase from $t_{Anr}$ to
$t_{tun}$.
In the single throat case, after the last KK mode becomes
non-relativistic at $t_{nr}$, the universe remains radiation
dominated. In the double throat case, radiation dominance is delayed
until $t_{tun}$.

The graviton production during this period can be estimated using
Eq.~(\ref{gravprod}). Similarly the integration is mainly contributed
from the early epoch around $t_0$, where $t_0$ is determined by the
initial condition $H_0^2
\approx \frac{1}{4t_0^2} \approx \frac{T_3 h_A^4}{3\mpl^2}$. We get
\bea
\rho_{A}^{grav} \approx g_s^{-1/4} 
\talpha^{1/4} h_A m_s \mpl^3 \sigma_{Kg} 
\frac{1}{t^2} ~.
\label{gravprodA}
\eea
The relative graviton density is
\bea
\Omega_{BBN~or~RDMD}^{grav} \equiv
\left( \frac{\rho_A^{grav}}{\rho_A^{tot}} \right)_{BBN~or~RDMD}
\approx g_s^{-1/4}
\talpha^{1/4} m_s \mpl \sigma_{Kg} h_A 
\left( \frac{t_{Anr}}{t_{tun}} \right)^{2/3} ~.
\label{gravratioA}
\eea
The last dilution factor is again due to the additional matter
dominated phase from $t_{Anr}$ to $t_{tun}$.
Using (\ref{csKg}) we see
that this is naturally a very small ratio.

\subsection{Tunneling from A-throat to S-throat}
\label{SecTunnelAS}

Because of the very small tunneling rate (\ref{trate1}) and
(\ref{trateAXbulk}), the tunneling of the KK modes in A-throat
effectively happens well after the temperature drops below the mass of
the lowest KK mode. Therefore the tunneling is dominated by these
non-relativistic lowest KK particles. 
The only way for the lowest KK mode to emit graviton is through
interaction with other particles. But
during the tunneling, they actually have long been decoupled from
each other and their kinetic energy has been largely red-shifted.  
(For this gas, we know that one can assign an effective
temperature $T \propto a^{-2} \propto t^{-4/3}$ if we keep a
correspondingly changing chemical potential.)
The energy density evolves similarly as in Eq.~(\ref{rhot}) 
except for a different numerical factor $4/3$.
To get the total graviton production, we integrate all particles as in
Eq.~(\ref{gravprod}).
Although the tunneling time scale tends to be very long as we see from
Sec.~\ref{SecKKGKP}, the graviton
production from the KK annihilation in this period is negligible
comparing to (\ref{gravratioA}), as we can see from a similar
integration as in Eq.~(\ref{gravprod}).
This is because the
spatial expansion keeps on red-shifting the KK density and velocity.

So the only constraint on the time scale of this period is that the
spatial expansion
during this long period does not cool the universe too much. This
requires that the energy density at the end of the tunneling,
$\rho_{KK}(t_{tun}) \approx \mpl^2 \Gamma_{tun}^2$, be larger than the
energy density during BBN, $(1~{\rm MeV})^4$. This gives a lower
bound on $h_A$. For example, for the resonance case or the
non-resonance case,
\bea
\Gamma_{tun}
\sim h_A^9 R_A^{-1} 
~~~{\rm or}~~~   
\Gamma_{tun} \sim h_A^{17} R_A^{-1} ~,
\eea
we need
\bea
h_A \gtrsim (R_A \cdot {\rm MeV}^2/\mpl)^{1/9}
~~~{\rm or}~~~
h_A \gtrsim (R_A \cdot {\rm MeV}^2/\mpl)^{1/17} ~,
\eea
respectively.
If for example $m_s \sim 10^{10} {\rm GeV}$ 
and $N_A^{1/4} \sim 10$, we need
\bea
h_A \gtrsim 10^{-4} ~~~{\rm or}~~~ h_A \gtrsim 10^{-2} ~,
\eea
respectively. 
If we want to incorporate TeV baryogenesis in the S-throat, then we
need
\bea
h_A \gtrsim 10^{-2} ~~~{\rm or}~~~ h_A \gtrsim 10^{-1} ~,
\eea
respectively.
These are reasonable for an A-throat. Note that for the
5-dim RS setup, the bound is much looser.

The KK modes tunneled into the S-throat will become thermalized in a
rate
\bea
\Gamma_{KK}(t_{tun}) \approx n(t_{tun}) \sigma_{KK}
\approx \left(h_S R_S^{-1} \right)^{-\frac{\gamma}{\gamma+4}} 
\left( \mpl \Gamma_{tun} \right)^{\frac{2(\gamma+3)}{\gamma+4}} 
\sigma_{KK} ~,
\eea
where the time dependence of the number density of the KK modes $n(t)$
is obtained from formula same as Eq.(\ref{gTA}) and
(\ref{TtA}). As long as this rate is larger than the tunneling rate,
which is similar to the Hubble expansion rate at the end of the
tunneling,
we can assume the thermal equilibrium. 
This requirement translates into a condition on $h_S$
which is easy to satisfy.\footnote{Ignoring all other
factors, this condition requires the warp factor $h_S <
\Gamma_{tun}^{\frac{\gamma+2}{5\gamma+16}} \cdot {\rm
mass}^{-\frac{\gamma+2}{5\gamma+16}}$. For example, 
according to the discussion in
Sec.~\ref{SecLong} and the result (\ref{hXcond}), this is naturally
satisfied.}

Since the warp factor of the S-throat is much smaller than that of the
A-throat, the KK spectrum is much denser. After the thermalization,
the KK modes will decay to the much lower mass levels and become
much
more difficult to tunnel back out. Therefore the tunneling from A to
S is effectively a one-way process. This streaming of the KK modes
heats
the S-throat. At the end of the tunneling process, the universe will
become dominated by the thermalized KK modes in the S-throat and
become radiation-dominated again.

But there is a subtlety here. Due to the same reason, i.e.~the
S-throat has a much smaller warp factor, the initial thermalized gas
may become stringy. To get an estimate on the temperature
using the KK spectrum (\ref{rhotTt}), after tunneling we have
\bea
\frac{\mpl^2}{t_{tun}^2} \approx 
\frac{T_S^{\gamma+4}}{(h_S R_S^{-1})^\gamma} ~.
\eea
Borrowing the result (\ref{hmincond}) from Sec.~\ref{SecMulti}, 
we see that 
\bea
T_S(t_{tun}) \lesssim 
h_S^{\frac{\gamma+2}{\gamma+4}} N_S^{\frac{1}{\gamma+4}} 
\mpl^{\frac{2}{\gamma+4}} R_S^{-\frac{\gamma+2}{\gamma+4}} 
\propto h_S^{\frac{\gamma+2}{\gamma+4}} m_s ~.
\label{TSbound}
\eea
On the one hand, we see that this upper bound is typically much
smaller than the 
mass level $h_A R_A^{-1}$ of the lowest KK mode in the A-throat; on
the other hand, we see that it is possible for $T_S(t_{tun})$ to
exceed the
red-shifted string scale $h_S m_s$ in the S-throat. 
To proceed, we will
ignore the involvement of the possible string states for the
following reasons. First, as universe cools down the thermalized gas
will eventually become a gas of KK particles. So the estimates on the
KK relics still apply. Second, in the most pedagogical sense, since
(\ref{TSbound}) is just a upper bound, we may adjust various
parameters to make $T_S(t_{tun})$ stay below the red-shifted string
scale. However, a better understanding of such a possible stringy phase
and its consequences
is clearly very interesting, because it may not happen too early away
from the BBN. 

Before continuing our story in the S-throat, we note that there is
another possible period during the tunneling that deserves
attention. As we have seen from Sec.~\ref{SecKKGKP}, if bulk resonance
happens, KK will stay in the bulk in this intermediate period, so it
is also important to analyze the graviton production and possible KK
relics in this case. It is easy to check that any KK in this resonance
period is decoupled from each other. So any stable KK, once produced
through decay process,
will be a problem. But this is not a big concern, because there is
usually no good candidates for stable KK modes in the bulk. The
wavefunction of these KK modes spread over the bulk, and will always
hit some throats. This is in fact how we estimated the tunneling rate
to X-throat in Sec.~\ref{SecBetween}. 
For graviton production, we integrate all particles and get a very
small ratio
\bea
\frac{\rho^{grav}}{\rho^{tot}} \approx
g^{1/4} \mpl^{3/2} \sigma_{Kg} \Gamma_{tun}^{1/2} ~,
\label{gravratioRes}
\eea
where $\Gamma_{tun} \sim h_A^9 R_A^{-1}$ is the tunneling rate from
the A-throat to the bulk. 

\subsection{KK evolution in S-throat}
\label{SecKKinS}

The evolution of KK modes in S-throat is similar to that in
A-throat. A small difference is the additional relativistic degrees of
freedom $g_0$ due to the Standard Model particles at the end of the
mass spectrum. So we take $g$ to be $n_{KK}^\gamma$ if $g>g_0$, and
approximately constant $g_0$ later on.

The relative abundance of the possible KK relics is similarly
obtained,
\bea
\left( \frac{\rho^{KK}_{dec}}{\rho^{tot}} \right)_{RDMD} 
\approx \sum_{n_{KK}} g^{-1/4} \mpl^{-1} \rho_{RDMD}^{-1/4}
\sigma_{KK}^{-1} ~.
\label{relicsS}
\eea
Comparing to (\ref{relicsA}) this is typically 
negligible since $\sigma_{KK}$
here is enhanced by powers of $h_S \ll h_A$.

The graviton is mainly produced around $t_{tun}$,
\bea
\left( \frac{\rho^{grav}}{\rho^{tot}} \right)_{BBN~or~RDMD} \approx
\talpha_S^{1/4} \left( h_S T_3^{1/4}
\right)^{-\frac{\gamma}{\gamma+4}}
\mpl^{\frac{2\gamma+6}{\gamma+4}} \Gamma^{\frac{\gamma+2}{\gamma+4}} 
\sigma_{Kg} ~.
\label{gravratioS}
\eea
For small $\gamma$, this ratio is much smaller than
(\ref{gravratioA}) due to the dilution of the KK particles 
during the tunneling.

Overall,
we summarize the graviton and KK abundance obtained in this
section and Sec.~\ref{SecKKprod} for the double throat case.

For the graviton, the total abundance is a summation of those from
the initial decay in Sec~\ref{SecKKprod}, $\left( \frac{R}{L}
\right)^6 h_A^2 \left
( \frac{t_{Anr}}{t_{tun}} \right)^{2/3}$ (where the last extra factor
is added due to the matter-dominated phase during the tunneling), the
annihilation in the
A-throat Eq.~(\ref{gravratioA}), the S-throat Eq.~(\ref{gravratioS}),
and the bulk Eq.~(\ref{gravratioRes}) 
if resonance happens.
All of them are suppressed by various combinations of warp factors,
bulk 
size and matter-dominated phase, so easily become very small. The
actual value is parameter dependent.

For the KK relics, Eq.~(\ref{relicsS}) is suppressed by powers of
the S-throat warp factor, so the A-throat contribution
(\ref{relicsA}), if stable, is typically larger. 
Evaluating (\ref{relicsA}) using
(\ref{csKK}) and (\ref{tAnr}), we get
\bea
\Omega^{KK}_{RDMD} \sim \left( \frac{R_A}{L} \right)^6 h_A
\mpl^{3/2} \rho_{RDMD}^{-1/4} R_A \Gamma_{tun}^{1/2} ~.
\label{OmegaKKinA}
\eea
For example, if $\mpl \approx 10^{18} ~{\rm GeV}$, $\rho_{RDMD}^{1/4}
\approx 1~{\rm eV}$, $m_s \approx 10^{10} ~{\rm GeV}$ and $N_A
\approx 10^4$, for the resonance case we need
\bea
\left( \frac{R_A}{L} \right)^6 h_A^{5.5} \lesssim 10^{-31}
\label{relicsAeg1}
\eea
to satisfy $\Omega^{KK}_{RDMD} \lesssim 1$. This for example can be
satisfied if $L\approx 10^3 R_A$ and $h_A \approx 10^{-3}$. Note that
the resonance condition typically requires a large $L$.
For the non-resonance case, we need
\bea
\left( \frac{R_A}{L} \right)^6 h_A^{9.5} \lesssim 10^{-31} ~.
\label{relicsAeg2}
\eea
This for example can be satisfied if $L \approx 10^2 R_A$ and
$h_A\approx 10^{-2}$.

\section{The bulk case}
\label{SecBulk}

Before going to the more general multi-throat case, let us look at the
case where either SM branes or brane annihilation are in the bulk.

So far we have considered SM branes in throats. These throats can be
regarded more generally as black hole like geometry, created by large
number of D3-branes, 3-form fluxes or even induced large D3-charges on
the D7-branes. To enter these geometries KK modes need to
tunnel. But it is also possible that some small number of branes
appears in the bulk rather than hiding in some throats. For these
branes, the warping they induce is small so KK modes do not have to
tunnel a significant barrier. In this situation, if the KK
wavefunction appears in the bulk, it
can intersect with these branes and the KK particles can decay into
light brane fields. So they start to heat the bulk branes instead of
heating a throat through tunneling. This can happen in the bulk
resonance case when KK propagate in the bulk, or in the non-resonance
case where KK also has a wavefunction in the bulk although it is not
propagating. Namely these small number of branes in the bulk may
become SM branes.

The heating in this section can be easily studied by
collecting a few results in Sec.~\ref{SecSingle} and \ref{SecDouble}.
There are three possibilities. 

We first discuss the case in which the
brane annihilation still happens in a throat (A-throat). 
The heating of the SM
branes is caused by the KK particles tunneled out of the A-throat. 
The analyses in the A-throat are the same as that in
Sec.\ref{SecKKinA}. The difference is that, the KK particles will
tunnel to the bulk and decay to lower KK modes and SM brane
fields. Same as we discussed at the end of Sec.\ref{SecTunnelAS}, 
any stable KK will be dangerous in this case, 
but the requirement of no stable KK
is reasonable for the bulk since they will always intersect with the
bulk SM branes and decay. The graviton production is small and the
same as (\ref{gravratioRes}). So the thermal history is
rather similar to the double throat resonance case and the heating works.

The second case is when both brane annihilation and SM branes 
are in the bulk. The
analyses for the KK modes are the same as in the single throat case.
We get the same expressions as in
Eqs.~(\ref{KKratio}) and (\ref{gravratio}). 
But in this case, there
is no longer warp factor and bulk size enhancement factors to the KK
interactions, $\sigma_{KK} \approx \mpl^{-2}$ and we get
\bea
\Omega_{RDMD}^{KK} \approx \sum_{n_{KK}} g^{-1/4} \mpl
\rho_{RDMD}^{-1/4} ~.
\eea
However we still do not worry about the
KK relics in the bulk, because they are unstable against decaying into
brane fields. 
The graviton production during the KK evolution is still very low. 
Overall, for this case, the main graviton
abundance problem comes from the initial brane and string decay period
discussed in Sec.\ref{SecKKprod}.  Namely, 
the analyses in Ref.~\cite{Chialva:2005zy} shows
that in flat space the graviton is produced more than KK particles in
the initial string decay. In all other cases that we
considered, the brane annihilation happens in a throat so this
graviton
problem is absent. This is the only case that can have a serious
problem in heating.
This problem is present in models such the branes-at-angle scenario
\cite{Garcia-Bellido:2001ky}, 
the D3-D7 inflation with D7-branes in the bulk \cite{Dasgupta:2002ew}, or
the D3-anti-D3-brane scenario in the bulk
\cite{Burgess:2001fx,Dvali:2001fw}, if the SM branes are also in the
bulk.

This graviton problem can be cured if there is a matter-dominated era 
before the BBN, so the graviton density can be considerably red-shifted.
This leads to the third case where we locate the SM branes in
an
S-throat, although the brane annihilation happens in the bulk. If there
is no extra small number of (anti)branes left in the bulk, the KK
modes will eventually tunnel into the S-throat. This introduces a
matter dominated phase similar to what we have seen in the double
throat case. The graviton problem is hence cured by the red-shift and
the story of the S-throat in Sec.~\ref{SecKKinS} applies here.

\section{Multi-throat heating}
\label{SecMulti}

In this section, we consider the heating in a more general
multi-throat compactification. In this case, the brane-anti-brane
annihilation at the end of the inflation still happens in the
A-throat. The heating from the A-throat to the bulk or the
S-throat follows the same analyses in the previous
section, but the presence of many other throats gives additional
interesting phenomena.

\subsection{Throat deformation in Hubble expansion}

The spectrum in a throat is mainly determined by its minimum warp
factor. However in the Hubble expansion background, the warping of a
throat cannot be arbitrarily small. Even if we start with an
infinitely long throat,
the Hubble energy will impose an
infrared cutoff on the warp factor.

This Hubble back-reaction may be analyzed in different ways
\cite{Buchel:2002wf,Chen:2005ad,Frey:2005jk,Giddings:2005ff}.  
One way
to look at it is to consider the back-reaction of the closed string
created in the IR side of the warp space \cite{Chen:2005ad}. 
Assuming an infinitely long
throat, such string creation will be inevitable because the local
string scale will be red-shifted down the throat and eventually fall
below the Hubble energy scale. The energy density of these strings is
$H^4$. The back-reaction of such closed
strings become significant to the source of the throat if
\bea
H^4 \approx h_{min}^4 N T_3 ~,
\eea
since the throat with $R^4 \approx N/T_3$ can be thought of being
sourced by $N$ number of branes. Therefore, although in static
background we can have multiple throats with arbitrary warp factors,
in the Hubble expansion background they will have a cutoff
at
\bea
h_{min} \approx \frac{H R}{\sqrt{N}} 
\label{hmin}
\eea
if they are too long.

Another way to look at it is to consider various moduli
that stabilize the throat \cite{Frey:2005jk}. 
Although in static background, throat with
arbitrary warp factors can be constructed by stabilizing those moduli,
the moduli mass will obtain a correction in a Hubble expansion
background, and therefore cause the deformation of the throat. The
explicit evaluation along this line has so far been difficult due to
the subtleties analyzing the supergravity equations in the presence of
small warp factors \cite{Frey:2005jk,Giddings:2005ff}.

Here we give a toy model of this type to roughly
estimate this effect. Consider an infinitely long throat generated by
$N$ stack of branes rather than fluxes. In the Hubble expansion
background, the brane will fluctuate in the transverse directions,
\bea
\Delta \phi \approx \sqrt{T_3} \Delta r \approx \frac{H}{2\pi} ~.
\eea
This fluctuation causes a deviation of the brane moduli which
otherwise are located exactly at the origin. So the throat is no
longer
extended to infinity at $r=0$ but terminated at a minimum $r_{min}
\approx \Delta r$. This
corresponds to a minimum warp factor
\bea
h_{min} \approx \frac{\Delta r}{R} \approx \frac{H R}{\sqrt{N}} ~.
\label{hmin2}
\eea
We see that in this toy model we get a consistent picture as in
Eq.~(\ref{hmin}). 
This consistency can be regarded as a duality relation between
deformed AdS and
CFT: Eq.~(\ref{hmin}) is obtained from the point of view of
the gravity side,
while Eq.~(\ref{hmin2}) the field theory side.

These analyses apply to both the inflationary and heating epoch,
and give a time-dependent minimum value for throat warp factor at
corresponding
moment. As a remark, we note that the energy associated with those
closed strings and moduli potentials is negligible to the heating
energy, since it is of order $\cO(H^4)$ and much less than the
inflationary energy. But it can create a spectrum of cosmic
strings with relatively low tension \cite{Chen:2005ad}.

\subsection{Selection of long throats}
\label{SecLong}

In the presence of many throats, the tunneling of KK modes can have
many channels. That is, it is possible for the KK modes to go from
A-throat to different throats. So which throats become dominantly
heated depends on the tunneling branching ratio.

A necessary condition for the tunneling discussed in
Sec.~\ref{SecBetween} and {\ref{SecRes} 
to happen is the matching of energy levels in
both throats. This condition requires that an energy eigenvalue in the
X-throat fall within the energy width of the initial state in the
A-throat. The width of the lowest KK mode in the A-throat is
determined by its tunneling rate $\Gamma_{tun}$, 
as one can see from the
uncertainty principle. So, the more finely spaced
is the X-throat spectrum, the more likely such a matching can be
found.\footnote{Note that this is different from the argument that the
large phase
volume in a long throat guarantees more heating after string decay
because of equipartition, as discussed in
Ref.~\cite{Barnaby:2004gg,Kofman:2005yz}. As pointed out in the same
references, the presence of the potential barrier invalidates such
arguments. Here even for a short
throat, if a mass level matching is accidentally satisfied, the
tunneling will still proceed (but may oscillate). Decreasing warp
factor increases this chance.} 
A long throat is thus favored during the tunneling
heating. Since in a throat with a minimum warp
factor $h_X$, the energy spacing for the s-wave is $h_X R_X^{-1}$, if
\bea
h_X \lesssim \Gamma_{tun} R_X ~,
\label{hXcond}
\eea
the tunneling will generically happen for this throat.
The Hubble constant during the tunneling is $H \approx
\frac{1}{t_{tun}} \approx \Gamma_{tun}$,
so the allowed minimum warp factor
(\ref{hmin}) in this Hubble expansion background is 
\bea
h_{min} \approx \frac{\Gamma_{tun} R_X}{\sqrt{N_X}} ~.
\label{hmincond}
\eea
So the condition (\ref{hXcond}) is allowed.

Thus we have seen a dynamical mechanism of selecting a long Standard
Model throat in the multi-throat heating. For example, for the
tunneling rate (\ref{trateAXbulk}), if roughly $R_A \sim R_X \sim D$, 
$\Gamma_{tun} \sim h_A^9 D^{-1}$, any throat with 
\bea
h_X \lesssim h_A^9 
\eea
will be generically heated, while heating probability for a
shorter throat is smaller. A throat with a large hierarchy ratio
($\sim 10^{-10}$) in
the sense of Randall and Sundrum can be naturally selected in this
way. If the SM branes are in the bulk, such a mechanism will favor a
large bulk size.

As a comment, we note that, even if the mass levels between the A and
X does not match, tunneling can proceed through a secondary effect. A
virtual KK particle is first created in the X-throat, and then splits
into several lower mass states such as lower KK modes or brane
fields. The kinetic energy of these light particles can compensate the
mass level difference. It will be interesting to estimate the
branching ratio of such a process and we leave it for future study.

\subsection{Warped KK dark matter and hidden dark matter}
\label{SecDM}

In this subsection we discuss two interesting possibilities of
non-relativistic dark matter. The first is the KK relics in the
S-throat. In our setup, the lowest KK mode in the S-throat will
interact with the SM branes and decay into brane fields, i.e., SM particles as light 
open string modes. 
However, higher level KK modes can have a long enough life
time, e.g., if they are associated with a conserved angular momentum in the
throat. For example, in the Klebanov-Strassler throat, 
besides the excitation quantum number $n$, the KK modes can also have  
angular quantum numbers $j$ for the approximate $S^3$ and $l$ for $S^2$ in 
a realistic throat, i.e., a warped deformed conifold. 
The (approximate) $S^3$ remains finite at the bottom of the throat (a property
of the deformation of the warped conifold). As a consequence, $j>1$  
quantum number costs less energy than the $l$ quantum numbers \cite{Firouzjahi:2005qs}. 
The KK modes with only $j$ quantum numbers have non-vanishing wave functions 
at the bottom of the throat, so they can easily decay to SM particles. 
(Recall that transverse momenta can be absorbed by the solitonic-like branes.)
On the other hand, the $S^2$ shrinks to zero at the bottom of the throat, so the 
wave functions of $l>0$ KK modes are suppressed at the origin, much like that 
in the hydrogen atom. Having no overlap with the SM branes, the light $l>0$ modes 
will not decay to SM particles. 
They couple to SM particles via the exchange of a s-wave KK mode, so two $l>0$ 
KK modes can annihilate.
However, this annihilation rate drops rapidly as the universe
expands, and these KK relics abundance can be estimated as we did for
the simplified throat in the previous sections.
Their tunneling rate is very small due to the
smallness of the S-throat warp factor and additional non-s-wave
suppression factors. So these lightest angular KK modes are quite stable.

Such a non-relativistic relic was a concern in
Ref.~\cite{Kofman:2005yz} since
their abundance was estimated to be extremely large. So if stable they
would over-close the universe. In the previous sections, we have
analyzed the KK thermal evolution and the abundance of these KK relics
more carefully. 
As we discussed, for example in Eq.~(\ref{thermalcond}), these relics
typically decouple non-relativistically.
From 
Eq.~(\ref{relics}) and (\ref{relicsS}), we can see
that the relative density of these KK relics is suppressed by warp
factors, bulk size and matter-dominated tunneling phase, 
and so can be naturally made safe. 
(To compare with the analysis in
Ref.~\cite{Kofman:2005yz}, we note Eq.~(4.16) in \cite{Kofman:2005yz}
is based on relativistic decoupling and has no adjustable small
parameters. Here Eq.~(\ref{relics}) and
(\ref{relicsS}) have several important differences: a factor of
$h_{A,S}^2$ due to the non-relativistic decoupling; 
a factor of $(R/L)^6$
from the throat-versus-bulk effect; and in the double throat case, an
extra suppression $(t_{Anr}/t_{tun})^{1/2}$ for the A-throat from the
matter-dominated
tunneling phase.) 

Of course, one can
then tune various parameters in the formulae to make them just
account for the density of the observed dark matter. Namely they can
become the warped KK dark matter.
In the single throat case, the relics abundance is given by
(\ref{relics}); in the double throat case, (\ref{relicsS}) is
typically smaller than (\ref{relicsA}), but it may become important if
the KK relics in A-throat have too short (tunneling or decay)
lifetime. (Otherwise we will discuss it in
Sec.\ref{SecRays}.)
They may be detected by collider experiments or underground dark
matter searches. For example, when two SM particles collide at the
red-shifted KK energy scale, they can
produce an intermediate unstable s-wave KK particles, then
subsequently
decay into two warped KK dark matter particles.
They will have different cross-sections than those of the more
standard dark matter candidates.
Discussions on various other kinds of KK dark
matter can be found in e.g.,
Ref.~\cite{Dimopoulos:2001ui,Dimopoulos:2001qd,Dienes:1998vg,Cheng:2002ej,Servant:2002aq}.

\medskip
Next we consider another type of dark matter.
In the multi-throat setup, when the KK particles are tunneling from
the A-throat to heat the S-throat or the bulk SM branes, 
similar process can also happen
for any throat which satisfies the mass level matching condition
discussed in Sec.~\ref{SecLong} and Sec.~\ref{SecKKGKP}. 
In particular, such a tunneling
happens generically for any throat long enough to satisfy
(\ref{hXcond}). 
Even in the single throat case, it is possible for a tiny amount of KK
modes to tunnel out to other throats.\footnote{
The possibility that KK particles may tunnel to other
throats is also discussed in
Ref.~\cite{Kofman:2005yz,Chialva:2005zy}.
Ref.~\cite{Kofman:2005yz} raised a concern
that this would generate KK relics that over-close the universe and
destroy the BBN. In Sec.~\ref{SecCoin}, we will see how the unusual
properties of the hidden dark matter can be used to avoid such
problems.
Ref.~\cite{Chialva:2005zy} gave a lower bound on these throats to
avoid the overheat of the S-throat
after BBN due to KK tunneling from those throats.}
In this subsection and Sec.~\ref{SecCoin}, we propose that the energy
density going into those throats can create new dark matter
candidates. Using the knowledge of the thermal evolution
that we obtained in this paper,
we will discuss the properties and thermal histories of these dark
matter candidates and the plausibility of this proposal, in particular
paying attention to the dark
matter coincidence problem.
We call those throats as D-throats, with D stands for
dark matter, and call those dark matter as hidden dark matter, 
because they are in a hidden sector and their wavefunction has
exponentially small overlap
with our visible sector.

The identities of the hidden dark matter in D-throats can be very
different from those of the warped KK dark matter. For a D-throat
without any branes in the end, as in Sec.~\ref{SecKKinA},
the KK tower will eventually be lowered to the
ground state. So the hidden dark matter in such a throat will be the
lowest KK particles, but not those with conserved angular
momentum. These KK particles are mutually decoupled and cannot
find each other to annihilate. Different from Sec.~\ref{SecKKinA},
since the D-throat has a
much smaller warp factor than the A-throat, 
these lowest KK modes take much
longer to tunnel out and therefore become stable.
For a D-throat with branes in the end, the identity of the
hidden dark matter can be either similar to that of the warped KK dark
matter in the S-throat, or more exotic particles analogous to our
baryons.

These dark matter have very different properties comparing to the
usual dark matter candidates such as the warped KK dark matter mentioned
above, the SM KK dark matter \cite{Dienes:1998vg,Cheng:2002ej,Servant:2002aq},
axions \cite{Peccei:1977ur,Wilczek:1977pj,Weinberg:1977ma}, or
the lightest supersymmetric particles
(LSP) \cite{Goldberg:1983nd,Ellis:1983ew}:

\begin{enumerate}
\item {\it Coupling:} 
For the usual dark matter candidate, although they are decoupled to
the Standard Model particles at low energy, they can couple if
the energy is excited reasonably higher. For example, for the warped KK
dark matter in the S-throat, the corresponding energy scale is the
warped KK scale, lower than the red-shifted string scale; for the LSP
in the MSSM, it is around the supersymmetry breaking scale TeV.

However, for the hidden dark matter in the D-throats, it almost
completely decouples from the Standard Model. As we can
see from Eq.~(\ref{wavetunnel1})-(\ref{wavetunnel3}), the
wavefunction of a KK mode damps at least by a factor of $h_D^8$ from D
to S-throat. The particles in the S-throat can couple to the hidden
dark matter only if they are excited out of the throat, namely by
roughly the UV string scale. Hence practically, the only interaction
between the hidden dark matter and the Standard Model (or its
extension) is through graviton mediation. Of course the hidden
dark matter can have self-interactions.

\item {\it Thermal history:}
For the usual dark matter, they initially thermalize with 
the rest of the
particles and closely participate in the thermal
history when the temperature in the S-throat is high. Only later on
after freeze-out, they evolve independently.

However, the hidden dark matter in D-throats has its own thermal
history. They have different temperatures once particles
thermalize in the D-throats. The thermal evolutions of the particles
in the D-throats and the S-throat are independent, with the net effect
driving the spatial expansion of the whole universe. 

\item {\it Relic density:} 
For the usual dark matter, the relic density is set by the
decoupling condition in the course of the thermal evolution. The
decoupling happens when the density of the particles drops low enough
so that their annihilation rate is below the expansion rate. After
that, they are driven apart forever by spatial expansion and cannot
find each other to annihilate. We have seen such an example in the
discussions of the KK relics in the previous sections.

But for the hidden dark matter in D-throats, it never evolves in the thermal
history of the S-throat and decouples at the beginning as soon as KK
particles tunnel to the D-throats and decay. 
The relics density is
set by the initial conditions during the tunneling and the properties
of the D-throats.

\end{enumerate}

\section{Hidden dark matter}
\label{SecHidden}

In the first subsection, we study more details of the hidden dark
matter in D-throats proposed in Sec.~\ref{SecDM} and the dark matter
coincidence problem.
In the second subsection, we discuss another type of 
hidden dark matter in the
A-throat and its possible implications on cosmic rays.

\subsection{Hidden dark matter and dark matter coincidence problem}
\label{SecCoin}

We first list some interesting consequences and constraints
followed from the properties discussed in Sec.~\ref{SecDM}. In this
subsection, when we make explicit comparisons, we will choose to compare
the hidden dark matter with the LSP dark matter for example.
Also we will choose S-throat as the location of SM branes. The
discussion still 
applies if we replace it with SM branes in the bulk.

\begin{enumerate}
\item
The mass of the LSP has to be greater than 100 GeV,
because otherwise it would have already shown up in colliders. But
this is not the concern for the hidden dark matter. In fact there is
little constraint on the mass of the hidden dark matter
particles. A more relevant statement is that the hidden dark matter
does not have to become non-relativistic 
before the temperature is 100 GeV in the
S-throat. It only has to take effect before the radiation-matter
transition epoch in the S-throat at around 1 eV.

\item
Each throat, including the S-throat and D-throats, has its own
RDMD transition epoch. To be consistent with our
observation, the S-throat has the latest RDMD
transition. 
The RDMD transition in D-throats can have different nature, depending
on their matter contents. We will give two examples later.
For an observer in a D-throat, after its local
RDMD transition, she will observe a period of hot dark
matter-dominated era, contributed by the S-throat.

\item
From the observations, we know that the density of the dark matter is
approximately 5 times of that of the baryons. If the hidden dark
matter in D-throats accounts for most of these dark
matter, we require
\bea
\rho_D(t_D^{RDMD}) \frac{a^3(t_D^{RDMD})} {a^3(t_S^{RDMD})} \approx 5
\rho_S(t_S^{RDMD}) ~,
\eea
where the subscript D or S refers to the local quantity of the D or
S-throat respectively.

Moreover, at the time of BBN, we require the density (of either matter
or radiation) in the D-throats be
much smaller than that in the S-throat, so that the process of the BBN
is not affected. Therefore during the tunneling heating, the most
heated throat will turn out to be the S-throat. This is consistent
with the previously mentioned requirement that the RDMD transition 
in the S-throat happens last. 
In fact, the energy distribution between the S and D-throats does not
have to be too different. The requirement
\bea
\frac{\rho_D(t_S^{BBN})} {\rho_S (t_S^{BBN})} \lesssim {\rm
a~few~percent}
\label{rhoDScond}
\eea
will be enough to be consistent with the observations.
This variation of the initial densities is due to the variation of the
tunneling rate with respect to different throats.
Indeed from the formulae (\ref{trate1}), (\ref{trateAXbulk}) and
(\ref{hXcond}), we can see
that, while the tunneling rate has the same scaling 
dependence on the warp
factor $h_A$, variations are expected for D-throats having different
locations, length scales or warp factors.

\end{enumerate}

We now look at two examples with different contents of hidden dark
matter in D-throats. 
We will focus on the following coincidence problem: why the
density of the dark matter happens to be roughly the same as (five
times) that of
the baryons? For the usual dark matter, this density is formed when
the density of the dark matter particles exponentially decreases and
eventually freezes out after the annihilation rate drops below the
Hubble expansion rate. Then this question amounts to showing whether
the required annihilation cross section is naturally obtained given a
model. For example, it has been argued that the WIMP may naturally
have this cross section. 
Here, since we have very different properties, it will be very
interesting to see what new aspects hidden dark matter offers to this
classic question.

$\bullet$ {\it Example 1:}
Imagine a D-throat without any brane located at the tip, so the hidden
dark matter will be the lowest KK mode with mass $m_D^{KK} \approx h_D
R_D^{-1}$. This is the case of hidden KK matter.

In this example, the RDMD transition in the D-throat happens when the
local temperature crosses $m_D^{KK}$ at $T_D(t_D^{RDMD}) \approx h_D
R_D^{-1}$. At this moment, the coincidence problem requires the
energy density in the D-throat to be five times of the baryons in the
S-throat. Since the total energy density is still dominated by
radiation in the S-throat until $t_S^{RDMD}$, we obtain the ratio of
the energy density between the D and S-throat at and before
$t_D^{RDMD}$,
\bea
\frac{\rho_D}{\rho_S} \sim 5\frac{T_S(t_S^{RDMD})} {T_S(t_D^{RDMD})}
~.
\label{rhoratio1}
\eea
Suppose initially we have (\ref{rhoDScond}),  
Eq.~(\ref{rhoratio1}) indicates
that 
\bea
t_S^{RDMD} > 10^4 ~t_D^{RDMD} ~.
\label{tStDcond1}
\eea
To get a more explicit requirement on the
D-throat, we use Eq.~(\ref{rhotTt}) and find initially
\bea
\frac{T_D}{T_S} \approx \left( \frac{\rho_D}{\rho_S}
\right)^{\frac{1}{\gamma+4}} \left( \frac{h_D R_D^{-1}}{h_S R_S^{-1}}
\right)^{\frac{\gamma}{\gamma+4}} ~.
\label{TDTS}
\eea
At the local RDMD transition in the D-throat,
\bea
T_D(t_D^{RDMD}) \approx h_D R_D^{-1} ~.
\label{TD}
\eea
At the same time, we know from the condition (\ref{rhoratio1}) and
$T_S(t_S^{RDMD}) \approx 1 {\rm eV}$ that
\bea
T_S(t_D^{RDMD}) \approx \frac{\rho_S}{\rho_D} \cdot 5~{\rm eV} ~.
\label{TS}
\eea
Plug Eq.~(\ref{TD}) and (\ref{TS}) into (\ref{TDTS}), we get
\bea
h_D \approx \left( \frac{\rho_S}{\rho_D} \right)^{\frac{\gamma+3}{4}}
\left( \frac{R_S}{R_D} \right)^{\frac{\gamma}{4}} 
\left( 5~{\rm eV} \cdot R_D \right)^{\frac{\gamma+4}{4}} 
h_S^{-\frac{\gamma}{4}} ~.
\label{ex1hDcond}
\eea
In this example, the coincidence problem results in a tuning of the
D-throat warp factor for any ratio $\rho_S/\rho_D$, as long as the
condition (\ref{tStDcond1}) is satisfied. For the bulk SM
branes, one simply replaces 
$\rho_S$ with $\rho_{bulk}$, $h_S^{-1} R_S$
with $L$ in Eq.~(\ref{ex1hDcond}).

$\bullet$ {\it Example 2:}
Perhaps the S-throat does not have to be special in the sense that
branes can also be located in other throats such as D-throats as
well. In each D-throat, there are different versions of the ``Standard
Model'' on branes. In such a case, the relevant hidden dark matter
content is no longer the lowest KK mode, it is different matter fields
on branes, analogous to our baryons. 
More importantly, the mass of these matter is no longer of
the local KK mass scale. The hidden dark matter
is massless at
the leading order comparing to this scale, just like the baryons in
our Standard Model throat. This is the case of hidden
matter.

Again we naturally assume that S-throat and D-throats have similar
but variable 
heating branching ratios (for example, if they all satisfy
(\ref{hXcond})). We
pick the most heated throat as the S-throat for reasons discussed
above. 
The thermal evolution of the D-throats are similar to the S-throat. As
the universe expands, the KK tower lowers and eventually transfer its
energy to the brane fields. The branes fields are mostly massless
particles, but there can be a tiny fraction of asymmetric ``baryons''
matter generated by ``baryogenesis''. Since radiation red-shifts faster
than matter, the ``baryon'' matter eventually dominates in the
D-throats. What follows is a period of ``hot dark matter''
dominated era contributed by the S-throat. After that the whole
universe becomes matter-dominated.

In this case, the coincidence
problem reduces to the similarity between the
``baryogenesis'' efficiencies on the ``Standard Models'' in
the D-throats and that on the Standard Model in the S-throat.
To give an example, we assume initially $\rho_S \sim 20 \rho_D$. If
the D-throat matter density is similar to ours so that $\rho_S^{\rm
matter} \sim 0.2 \rho_D^{\rm matter}$, the RDMD transition happens
a little earlier in the D-throat -- by a factor of 100 in terms of the
scale factor.
That is, Eq.~(\ref{tStDcond1}) still applies here. By
the time that the RDMD transition happens in the S-throat, 
we have both the baryons and
hidden matter with similar abundance.
Both the mass of the hidden dark matter and its final abundance become
much more independent of the D-throat warp factor.

Both examples give very interesting illustrations of how hidden dark
matter can behave differently against the usual intuition. Naively, if
the energy density that goes into a dark matter candidate (e.g.~LSP)
is as large
as a few percent of that goes into the SM particles, it surely
over-closes the universe and destroys the BBN. This is because the
LSP dark matter candidates become non-relativistic before the
temperature is 100
GeV. In terms of time, this happens more than $10^{22}$ times earlier
than the RDMD
transition. Only extremely tiny fraction of energy is allowed to
deposit into the dark matter to account for such a large
red-shift. But hidden dark matter can become non-relativistic much
later, as late as in (\ref{tStDcond1}) -- reducing $10^{22}$ to $10^4$
-- which greatly shortens the
red-shift. This is achieved, in the first example, by having a
long D-throat and therefore very low KK levels; in the second example,
by having similar amount of matter and radiation between the hidden
sector and the visible sector.

\subsection{Hidden dark matter and cosmic rays}
\label{SecRays}

So far in this section, 
we have studied the properties of the hidden dark matter lying in
D-throats which are different from both S and A-throat. But in the
double throat case or the bulk case, the KK modes
with specific conserved angular momentum in the A-throat can also
become viable dark matter candidates. It is possible that these KK
modes have very small tunneling rate. For example, the
s-wave in the A-throat have a tunneling rate $h_A^9 R_A^{-1}$ or 
$h_A^{17} R_A^{-1}$. 
This s-wave is responsible for the tunneling heating. For a
$l$-th partial wave, the tunneling rate is expected to be 
reduced by a factor of
$h_A^{4l}$ for the simplified geometry studied in Sec.~\ref{SecKKGKP}
(same as the absorption rate in \cite{Klebanov:1997kc}). 
Hence a reasonable higher partial wave can be stable
against the tunneling. If this angular 
momentum is also approximately conserved
(now it can be associated with either $S^3$ or $S^{2}$ in the KS
throat since we do not have branes left in the A-throat), it
becomes a candidate of stable relics.
We have estimated the relics abundance of such KK modes in
Eq.~(\ref{relicsA}). We see that it again contains various tunable small
parameters and can be a viable dark matter candidate.

An interesting feature of this dark matter is that its properties
lie in between the hidden dark matter in D-throats and the warped
KK dark matter in S-throat. In terms of couplings, it is similar to
the former --- it only couples to
the visible sector through gravity and usually cannot be produced in
colliders. So it is still the hidden dark matter. But in terms of the
properties of the thermal history, it is more like the latter --- it
decouples as a non-relativistic thermal relics and its abundance can
be calculated in the usual way.

We have been familiar with the fact that large cosmic strings can be
left in the A-throat as a result of closed string evolution.
For these cosmic strings, they cannot tunnel.
The evolution of cosmic strings will lead to a scale-invariant spectrum (i.e.,  
cosmic strings will
come in all sizes in some distribution). In general, that will yield  
a density depending
on string tension $\mu$. It goes like $G_N \mu \Gamma$ of the  
critical density, where $\Gamma$ is a numerical number that is around  
10 to 100 \cite{Tye:2005fn}. With $G_N \mu < 10^{-6}$, large
cosmic strings contributes very little to the critical density.
But now we see that it is also possible that the closed string and KK
evolution lead to some stable KK modes in the A-throat. 
Then those KK modes can have a sizable density to contribute to the
dark matter.

Such a dark matter is typically very heavy. Take the double throat
non-resonance case as an example. In the numerical example
of Eq.~(\ref{relicsAeg2}), the mass of this hidden dark matter is
\bea
M_{DM} \sim 10^7 {\rm GeV} ~.
\label{DMAmass1}
\eea
One can easily change the string scale since the bound
$\Omega^{KK}_{RDMD} \lesssim 1$ on
(\ref{OmegaKKinA}) is not very sensitive to this variation. For
example, taking $m_s \approx 10^{13} {\rm GeV}$, $N_A \approx 10^2$ 
and still keeping the
tunneling rate $\Gamma_{tun} \sim h_A^{17} R_A^{-1}$ and the
other parameters same, we get $h_A \sim 10^{-2}$ and $L\sim 60 R$ to
saturate the bound. In this example, the mass of the hidden dark
matter becomes
\bea
M_{DM} \sim 10^{11} {\rm GeV} ~.
\label{DMAmass2}
\eea
There are many interesting consequences of such a dark matter.

First, we have seen that this hidden dark matter belongs to thermal
relics, so both (\ref{DMAmass1}) and (\ref{DMAmass2}) 
evade a well-known bound \cite{Griest:1989wd} 
which states that the maximum
mass of a dark matter is a few hundred TeV if it is a thermal relic.
This bound comes because larger dark matter mass leads to smaller
annihilation cross section and more red-shift, 
and therefore excessive dark matter density. 
The reason that we evade it 
is the extra matter dominated tunneling phase discussed in
Sec.~\ref{SecKKinA} and \ref{SecTunnelAS}. During this phase the total
energy of the universe, including that of this dark matter, is hung up
in the A-throat in a long matter-dominated era. The
radiation-dominated phase in the S-throat is therefore much
delayed. This
greatly increases the allowed ratio between the density of the dark
matter and the rest.

Second, this dark matter can have a lifetime longer than the age of
the universe, but occasionally tunnel to the visible S-throat. Namely
this becomes a novel example of an ultra heavy thermal relic being the
source of the ultra high energy cosmic rays. To estimate the lifetime
of those angular KK modes in the A-throat, we use the ansatz
$\tau \sim h_A^{-17-4l} R_A$.
In the above example (\ref{DMAmass2}), we can see that
the lifetime is longer than the age of the universe $10^{17}$s if
$l>2$.

Third, since this cosmic ray is due to the decay of the dark matter in
our galaxy, they do not have to travel a long distance between
galaxies to lose energy through interactions with CMB.
So the GZK cutoff \cite{Greisen:1966jv,Zatsepin:1966jv} 
does not apply. The energy of
these ultra high energy cosmic rays can be as large as $10^{20} {\rm
eV}$ such as that in Eq.~(\ref{DMAmass2}).
Since the source of such decay is
not local objects, the distribution of the cosmic rays is more
isotropic. 
%It will follow the density distribution
%$\rho_{DM}({\bf x})$ of the dark matter in our galaxy. 
Comparing to the
distribution of the dark matter annihilation, which is proportional to
$\rho_{DM}^2({\bf x})$, these cosmic rays will have a distribution
proportional to $\rho_{DM}({\bf x})$. Moreover, the energy of the
cosmic rays $M_{DM}$ correlates with the inflation scale $V_{inf}$ 
and the tension of the cosmic strings $\mu_F$,
\bea
M_{DM} \approx n_{KK} (g_s N_A)^{-1/4} \mu_F^{1/2} \approx
n_{KK} N_A^{-1/4} V_{inf}^{1/4} ~.
\eea
If $N_A^{1/4}$ is not too big, they are all close to each other 
in terms of order of magnitude. So in this scenario 
a measurement of the energy of the relevant cosmic rays predicts the
primordial gravitational wave and the cosmic string tension.

Fourth, the nature of this decay can be extremely stringy. This
happens
in the double throat case. Although the
original KK mode in the A-throat is a field theory mode, after it
tunnels to the S-throat, its mass is way above the local red-shifted
string scale due to the large warping of the S-throat. So detailed
observations of such decays may become a useful laboratory to compare
with the theoretical calculations in string theory.

\subsection{Comparison to other dark matters}
\label{GIMPHDM}

Besides the LSP that we have discussed, in this subsection, we give a
couple of more comparisons between other dark matter candidates and
the two main categories of dark matter candidates proposed here, 
i.e.~the 
warped KK dark matter and the hidden dark matter.

The warped KK dark matter shares
common features of most usual dark matter candidates, 
such as their thermal
properties and coupling to SM particles. For example,
conceptually it is similar to that proposed in the 5-d RS geometry
\cite{Dimopoulos:2001ui,Dimopoulos:2001qd}. Besides the differences
caused by those between the 5-d RS and 10-d GKP setup, 
there is no graviton KK in
5-d stable against decaying on the IR branes, so in
\cite{Dimopoulos:2001ui,Dimopoulos:2001qd}, such a stable 
particle was put in
by hand (``bulky''); here we have a natural and viable realization of
their ``bulky''.

The hidden dark matter is more unusual.
Our definition of the hidden dark matter is that they are in a hidden
sector, and their wave-function only have an exponentially small
overlap with our visible sector. As far as detailed properties are
concerned, they may have overlaps with various other kinds of dark
matter candidates. But they also have differences. For example,
superWIMPs or GIMPs (super-weakly or gravitationally interacting
massive particles) such as gravitino or graviton KK modes in
universal extra dimension (UED) also have gravitationally suppressed
interaction with other particles \cite{Feng:2003xh}. 
But these particles, for example the
gravitino, can be the decay product of a supersymmetric Standard Model
particles with a mass such as 1 TeV. Hence superWIMPs actually
couple stronger than the hidden dark matter (but weaker than the
warped KK dark matter in the S-throat). For example, if one
collides two SM particles at 1 TeV, they can become an intermediate
supersymmetric particles, through a relatively strong coupling, and
then decay to gravitino, through gravitationally weak coupling. 
But for hidden
dark matter, even if the energy of the colliding particles is of order
their mass level, they still can only be created through the mediation
of gravitons. Because of the same reason, the gravitino can inherit
the
usual relics thermal abundance of the mother particle, but hidden dark
matter in the D-throats has a completely independent thermal history.
As another example, there is a proposal of having ultra high
energy cosmic rays as decay products of non-thermal heavy dark matter
\cite{Chung:1998ua}.
But as we have discussed, our heavy hidden dark matter in the A-throat
is thermal relics, while the non-thermal hidden dark matter 
in the D-throats does not
have to be heavy at all. Of course, more detailed study and comparison
with various dark matter candidates is necessary and interesting.

\section{Conclusions and discussions}
\label{SecDis}

In this paper, we have studied the properties of the graviton KK modes
in the single and multi-throat compactification, including their
wavefunctions, tunneling properties and interactions.
We have studied the thermal history after the brane annihilation at
the end of the brane inflation, to test whether the heating 
is compatible with
the standard big bang nucleosynthesis. We have found that, as long as
the brane annihilation and the SM branes are not both in the bulk, 
the heating process is viable no matter where they are: one of them 
in the bulk,
in the same throat or in different throats, 
although different scenarios have different thermal histories.
The KK thermal relics and the graviton abundance are suppressed by
various small factors: warp factors, throat versus bulk size, and the
red-shifting factor during the tunneling period. The long tunneling
phase required in some
cases is not only compatible with a
viable heating, but also gives rise to several interesting phenomena:
an extra suppression for the KK relics, a dynamical
mechanism of selecting a long Standard Model throat, and the generation of
hidden dark matter. 
Our results on the KK thermal history and relics abundance show that
there are three types of new dark matter candidates: the warped KK
particles with specific angular momentum in the S-throat; the hidden
dark matter in D-throats; and the hidden KK particles with higher
partial wave in the A-throat. Physically they differ from 
each other in terms of either the relics thermal history or couplings 
to the visible sector.
The novel properties of the hidden dark matter have
especially interesting implications on the dark matter coincidence
problem and ultra high energy cosmic rays.

We have seen several possibilities that stringy excitations are
produced: in the A-throat during the initial brane decay, in the
S-throat after the tunneling from A to S, in any throat with too small
warp factors. In the first two cases, the energy density in the stringy
excitations can dominate the total energy density initially and then
lose it to lower KK modes. In the last case, it never
dominates. It will be interesting and challenging to understand if
there is any novel effects associated with these phases, besides the
cosmic strings discussed in 
Ref.~\cite{Jones:2002cv,Sarangi:2002yt,Copeland:2003bj,Polchinski:2004ia}.
For example, it is entirely possible to produce some relatively long
lived massive open string modes in the SM branes.
These modes will evolve as $a^{-3}$ for a while, eventually decaying
to the light SM particles. They tend to enhance the SM entropy
contribution during BBN. However, without an explicit model, it is
difficult to assess the importance of this possibility. Instead, we
see that the graviton
KK modes play that role beautifully, but in a way that is more intricate. Here, the 
important role played by the warped geometry cannot be over-emphasized.

The graviton KK modes in different throats are prime candidates
as dark matter.
The stable KK modes in the S-throat have qualitatively similar properties as usual dark matter candidates, but with different production and scattering cross-sections.
On the other hand, the KK modes in the D-throat or A-throat away from
the SM branes have rather
different properties. 
Since they do not interact with SM particles or their superpartners except via gravity, collider experiments and/or underground dark matter searches are useless. 
One way to detect the hidden dark matter is via gravitational lensing \cite{Wittman:2000tc}. 
With better CMBR data, one may fix the parameters in brane inflation more precisely. Together with structure formation simulations, rotational curves etc., they should allow us to better determine the KK spectrum and interactions, their production and subsequent evolution
as well as properties of the D or A-throat. One can then check to see if their distribution in our universe matches the lensing observations expected in the near future. 

It is quite possible that the hidden dark matter is only a component of the total dark matter expected. 
If direct detection of dark matter candidates and their subsequent measurements cannot
account for all the dark matter expected, this may be taken as an indirect evidence of 
hidden dark matter. Violation of the GZK bound on cosmic rays can be a different piece of evidence.

The UV DBI inflation \cite{Silverstein:2003hf,Alishahiha:2004eh} and
some special case of the IR DBI inflation
\cite{Chen:2004gc,Chen:2005ad}
require relativistic brane collision as heating, in addition to or as
an alternative to the brane annihilation. While we expect the main
thermal history after thermalization to be similar to one of our cases here,
it is interesting to see how the details may differ especially at
the initial moment \cite{McAllister:2004gd}.

In this paper, we try to give an overall picture of heating towards the end 
of brane inflation. It is an interesting story all by itself. In doing so, we have 
made only order-of-magnitude estimates and neglected all 
the numerical factors. A more precise analysis will probably increase
the likeliness of some scenarios and decrease that of others, so 
many interesting aspects are worthy to be studied in more details. 
It is clear that we have considered only some of the simplest scenarios.
There are variations of the scenarios considered here that deserve investigations.

\acknowledgments 
We like to thank Jim Cline, Pier Stefano Corasaniti, 
Hassan Firouzjahi, Min-xin Huang, Lam Hui, Lev Kofman, 
Peter Langfelder,
Louis Leblond, Kaya Mori, Maxim Perelstein, Sash Sarangi, Sarah
Shandera,
Jan Pieter van der Schaar, Gary Shiu, Bret 
Underwood, Ira Wasserman, Erick Weinberg and Haitao Yu
for very help discussions. 
We thank the center of mathematical science of Zhejiang University for
the hospitality.
XC thanks the Newman Laboratory
for Elementary Particle Physics in Cornell
University and Aspen center for physics 
for the hospitality. The work of XC is
supported in part by the U.S. Department of Energy.
The work of SHT is supported by the National Science Foundation under
grant PHY-0355005.

\appendix
\section{a KK cannot decay to all gravitons}
\label{AppKKdecay}

In this appendix, we give a general argument that a KK particle cannot
decay to all gravitons, generalizing that in Sec.~\ref{SecKKinter}.

The 10-dim
metric contains a warped four large dimensions and six
internal dimensions,
\bea
ds_{10}^2 &=& g_{\mu\nu}(x,y) ~dx^\mu dx^\nu + g_{mn}(y) ~dy^m dy^n
\nonumber \\
&=& A^{-1/2}(y) ~\hat g_{\mu\nu}(x,y)~dx^\mu dx^\nu 
+ g_{mn}(y) ~dy^m dy^n ~.
\label{ds10}
\eea
The warp factor $A^{-1/2}(y)$ and the internal metric $g_{mn}(y)$ is
generated by the matter such as fluxes and branes.
The wave-functions of KK modes have non-trivial dependence on
the internal coordinates, $\hat g_{\mu\nu} = \eta_{\mu\nu} +
h^{KK}_{\mu\nu}(x,y)$, while that of the graviton does not
depend on the internal coordinates, 
$\hat g_{\mu\nu} = \eta_{\mu\nu} +
h^{grav}_{\mu\nu}(x)$.
The 10-d action is 
\bea
S_{10} = M_{(10)}^8 \int d^{10}X \sqrt{-g_{(10)}} R_{(10)} + S_{\rm
matter} ~.
\label{S10}
\eea

To get the n-point coupling, we perturbatively expand the action
(\ref{S10}) to the polynomial of
$h_{\mu\nu}$ and $\partial_\rho h_{\mu\nu}$.
To examine the amplitude of one KK mode coupled 
to any number of gravitons,
we replace one of $h_{\mu\nu}$ with the
wave-function of the KK mode in all possible way in each term. 
We then factor out
the wave-function $A^{-1/2} h_{\mu\nu}^{KK}$, 
integrating by parts if necessary. 
This procedure is the same
as what we do to get the equation of motion, except that we now
replace $\delta g_{\mu\nu}$ with $A^{-1/2} h^{KK}_{\mu\nu}$.
So we get
\bea
S_{10} \supset - M_{(10)}^8 \int d^{10}X \sqrt{-g_{(10)}}
\left( A^{-1/2} h_{KK}^{\mu\nu} \right)
\left( R^{(10)}_{\mu\nu} - \half g_{\mu\nu}R^{(10)} -
\frac{1}{M_{(10)}^8}
T^{\rm matter}_{\mu\nu} \right) ~.
\label{KKgrav}
\eea
Under the metric (\ref{ds10}), 
\bea
R^{(10)}_{\mu\nu} = 
R^{(4)}_{\mu\nu} -\half \nabla_{(6)}^2 g_{\mu\nu}
-\frac{1}{4} g^{\rho\lambda} \partial_m g_{\mu\nu} \partial^m
g_{\rho\lambda} + \half g^{\rho\lambda} \partial_m g_{\nu\rho}
\partial^m g_{\mu\lambda} ~,
\label{Rmunu}
\eea
where $R_{\mu\nu}^{(4)}$ is constructed from 4-d metric $g_{\mu\nu}$, and
$\nabla_{(6)}$ is constructed from the internal metric $g_{mn}$. 
For the background warped geometry, the source does not depend on the
large dimensions,
\bea
\frac{\partial \CL^{\rm matter}} {\partial g^{\mu\nu}} =0 ~.
\eea
Therefore, $T^{\rm matter}_{\mu\nu}$ is proportional to $g_{\mu\nu}$,
\bea
T^{\rm matter}_{\mu\nu} = -\half g_{\mu\nu} \CL^{\rm matter} ~.
\eea
Except for the one KK mode, all the other $\hat g_{\mu\nu}$ in
(\ref{KKgrav})
are gravitons and therefore $y$-independent. Hence in the last three
terms of (\ref{Rmunu}), $\hat g_{\mu\nu}$ factors out, and we get
\bea
&& R^{(10)}_{\mu\nu} - \half g_{\mu\nu}R^{(10)} - \frac{1}{M_{(10)}^8}
T^{\rm matter}_{\mu\nu}  \nonumber \\
&=& \hat R^{(4)}_{\mu\nu} - \half \hat g_{\mu\nu} \hat R^{(4)}
\nonumber \\
&-& \hat g_{\mu\nu} \left( -\half \nabla_{(6)}^2 A^{-1/2} - \half
A^{1/2} \partial_m A^{-1/2} \partial^m A^{-1/2} + \half A^{-1/2}
R_{(6)} \right) 
\nonumber \\
&+& \hat g_{\mu\nu} \frac{1}{2M_{(10)}^8} A^{-1/2} \CL^{\rm matter} ~,
\label{eomFactor}
\eea
where $\hat R^{(4)}_{\mu\nu}$ is constructed from $\hat g_{\mu\nu}$.

The last two terms in (\ref{eomFactor}) cancel because the background
equations of motion are just these two terms with the overall factor 
$\hat g_{\mu\nu}$
replaced by $\eta_{\mu\nu}$, the rest does not depend on $\hat
g_{\mu\nu}$. The first two terms also cancel because
the graviton wave-function satisfies 
$\hat R^{(4)}_{\mu\nu} - \half \hat g_{\mu\nu} \hat R^{(4)} =0$ by
definition, which is beyond the linear approximation and includes all
possible non-linear couplings. Therefore (\ref{KKgrav}) vanishes
identically, namely, the decay product of one KK mode cannot be all
gravitons.


\clearpage


\begin{thebibliography}{}



%\cite{Dvali:1998pa}
\bibitem{Dvali:1998pa}
G.~R.~Dvali and S.-H.~H.~Tye,
``Brane inflation,''
Phys.\ Lett.\ B {\bf 450}, 72 (1999)
[arXiv:hep-ph/9812483].
%%CITATION = HEP-PH 9812483;%%


%\cite{Burgess:2001fx}
\bibitem{Burgess:2001fx}
C.~P.~Burgess, M.~Majumdar, D.~Nolte, F.~Quevedo, G.~Rajesh and R.~J.~Zhang,
``The inflationary brane-antibrane universe,''
JHEP {\bf 0107}, 047 (2001)
[arXiv:hep-th/0105204].
%%CITATION = HEP-TH 0105204;%%


%\cite{Dvali:2001fw}
\bibitem{Dvali:2001fw}
G.~R.~Dvali, Q.~Shafi and S.~Solganik,
``D-brane inflation,''
arXiv:hep-th/0105203.
%%CITATION = HEP-TH 0105203;%%


%\cite{Giddings:2001yu}
\bibitem{Giddings:2001yu}
S.~B.~Giddings, S.~Kachru and J.~Polchinski,
``Hierarchies from fluxes in string compactifications,''
Phys.\ Rev.\ D {\bf 66}, 106006 (2002)
[arXiv:hep-th/0105097].
%%CITATION = HEP-TH 0105097;%%

%\cite{Kachru:2003aw}
\bibitem{Kachru:2003aw}
S.~Kachru, R.~Kallosh, A.~Linde and S.~P.~Trivedi,
``De Sitter vacua in string theory,''
Phys.\ Rev.\ D {\bf 68}, 046005 (2003)
[arXiv:hep-th/0301240].
%%CITATION = HEP-TH 0301240;%%

%\cite{Grana:2005jc}
\bibitem{Grana:2005jc}
  M.~Grana,
  ``Flux compactifications in string theory: A comprehensive review,''
  Phys.\ Rept.\  {\bf 423}, 91 (2006)
  [arXiv:hep-th/0509003].
  %%CITATION = HEP-TH 0509003;%%

%\cite{Kachru:2003sx}
\bibitem{Kachru:2003sx}
S.~Kachru, R.~Kallosh, A.~Linde, J.~Maldacena, L.~McAllister and S.~P.~Trivedi,
``Towards inflation in string theory,''
JCAP {\bf 0310}, 013 (2003)
[arXiv:hep-th/0308055].
%%CITATION = HEP-TH 0308055;%%


%\cite{Firouzjahi:2005dh}
\bibitem{Firouzjahi:2005dh}
  H.~Firouzjahi and S.-H.~H. Tye,
  ``Brane inflation and cosmic string tension in superstring theory,''
  JCAP {\bf 0503}, 009 (2005)
  [arXiv:hep-th/0501099].
  %%CITATION = HEP-TH 0501099;%%


%\cite{Silverstein:2003hf}
\bibitem{Silverstein:2003hf}
  E.~Silverstein and D.~Tong,
  ``Scalar speed limits and cosmology: Acceleration from D-cceleration,''
  Phys.\ Rev.\ D {\bf 70}, 103505 (2004)
  [arXiv:hep-th/0310221].
  %%CITATION = HEP-TH 0310221;%%


%\cite{Alishahiha:2004eh}
\bibitem{Alishahiha:2004eh}
  M.~Alishahiha, E.~Silverstein and D.~Tong,
  ``DBI in the sky,''
  Phys.\ Rev.\ D {\bf 70}, 123505 (2004)
  [arXiv:hep-th/0404084].
  %%CITATION = HEP-TH 0404084;%%


%\cite{Chen:2004gc}
\bibitem{Chen:2004gc}
  X.~G.~Chen,
  ``Multi-throat brane inflation,''
  Phys.\ Rev.\ D {\bf 71}, 063506 (2005)
  [arXiv:hep-th/0408084].
  %%CITATION = HEP-TH 0408084;%%


%\cite{Chen:2005ad}
\bibitem{Chen:2005ad}
  X.~G.~Chen,
  ``Inflation from warped space,''
  JHEP {\bf 0508}, 045 (2005)
  [arXiv:hep-th/0501184].
  %%CITATION = HEP-TH 0501184;%%


%\cite{Shandera:2006ax}
\bibitem{Shandera:2006ax}
  S.~E.~Shandera and S.-H.~H. Tye,
  ``Observing brane inflation,''
  arXiv:hep-th/0601099.
  %%CITATION = HEP-TH 0601099;%%


%\cite{Kachru:2002gs}
\bibitem{Kachru:2002gs}
S.~Kachru, J.~Pearson and H.~Verlinde,
``Brane/flux annihilation and the string dual of a non-supersymmetric  field
theory,''
JHEP {\bf 0206}, 021 (2002)
[arXiv:hep-th/0112197].
%%CITATION = HEP-TH 0112197;%%


%\cite{DeWolfe:2004qx}
\bibitem{DeWolfe:2004qx}
O.~DeWolfe, S.~Kachru and H.~Verlinde,
``The giant inflaton,''
JHEP {\bf 0405}, 017 (2004)
[arXiv:hep-th/0403123].
%%CITATION = HEP-TH 0403123;%%

\bibitem{Cline:2005ty}
  J.~M.~Cline and H.~Stoica,
  ``Multibrane inflation and dynamical flattening of the inflaton potential,''
  Phys.\ Rev.\ D {\bf 72}, 126004 (2005)
  [arXiv:hep-th/0508029].
  
%\cite{Randall:1999ee}
\bibitem{Randall:1999ee}
  L.~Randall and R.~Sundrum,
  ``A large mass hierarchy from a small extra dimension,''
  Phys.\ Rev.\ Lett.\  {\bf 83}, 3370 (1999)
  [arXiv:hep-ph/9905221].
  %%CITATION = HEP-PH 9905221;%%


%\cite{Chen:2005fe}
\bibitem{Chen:2005fe}
  X.~G.~Chen,
  ``Running non-Gaussianities in DBI inflation,''
  Phys.\ Rev.\ D {\bf 72}, 123518 (2005)
  [arXiv:astro-ph/0507053].
  %%CITATION = ASTRO-PH 0507053;%%

%\cite{Jones:2002cv}
\bibitem{Jones:2002cv}
N.~Jones, H.~Stoica and S.-H.~H.~Tye,
``Brane interaction as the origin of inflation,''
JHEP {\bf 0207}, 051 (2002)
[arXiv:hep-th/0203163].
%%CITATION = HEP-TH 0203163;%%

%\cite{Sarangi:2002yt}
\bibitem{Sarangi:2002yt}
S.~Sarangi and S.-H.~H.~Tye,
``Cosmic string production towards the end of brane inflation,''
Phys.\ Lett.\ B {\bf 536}, 185 (2002)
[arXiv:hep-th/0204074].
%%CITATION = HEP-TH 0204074;%%

%\cite{Copeland:2003bj}
\bibitem{Copeland:2003bj}
E.~J.~Copeland, R.~C.~Myers and J.~Polchinski,
``Cosmic F- and D-strings,''
JHEP {\bf 0406}, 013 (2004)
[arXiv:hep-th/0312067].
%%CITATION = HEP-TH 0312067;%%

%\cite{Polchinski:2004ia}
\bibitem{Polchinski:2004ia}
J.~Polchinski,
``Introduction to cosmic F- and D-strings,''
arXiv:hep-th/0412244.
%%CITATION = HEP-TH 0412244;%%


%\cite{Sen:2002nu}
\bibitem{Sen:2002nu}
  A.~Sen,
  ``Rolling tachyon,''
  JHEP {\bf 0204}, 048 (2002)
  [arXiv:hep-th/0203211].
  %%CITATION = HEP-TH 0203211;%%

%\cite{Sen:2002in}
\bibitem{Sen:2002in}
  A.~Sen,
  ``Tachyon matter,''
  JHEP {\bf 0207}, 065 (2002)
  [arXiv:hep-th/0203265].
  %%CITATION = HEP-TH 0203265;%%

%\cite{Sen:2004nf}
\bibitem{Sen:2004nf}
  A.~Sen,
  ``Tachyon dynamics in open string theory,''
  Int.\ J.\ Mod.\ Phys.\ A {\bf 20}, 5513 (2005)
  [arXiv:hep-th/0410103].
  %%CITATION = HEP-TH 0410103;%%

\bibitem{Shiu:2002xp}
  G.~Shiu, S.-H.~H.~Tye and I.~Wasserman,
  ``Rolling tachyon in brane world cosmology from superstring field theory,''
  Phys.\ Rev.\ D {\bf 67}, 083517 (2003)
  [arXiv:hep-th/0207119].
  
\bibitem{Cline:2002it}
  J.~M.~Cline, H.~Firouzjahi and P.~Martineau,
  ``Reheating from tachyon condensation,''
  JHEP {\bf 0211}, 041 (2002)
  [arXiv:hep-th/0207156].

%\cite{Lambert:2003zr}
\bibitem{Lambert:2003zr}
  N.~Lambert, H.~Liu and J.~Maldacena,
  ``Closed strings from decaying D-branes,''
  arXiv:hep-th/0303139.
  %%CITATION = HEP-TH 0303139;%%


%\cite{Chen:2003xq}
\bibitem{Chen:2003xq}
  X.~G.~Chen,
  ``One-loop evolution in rolling tachyon,''
  Phys.\ Rev.\ D {\bf 70}, 086001 (2004)
  [arXiv:hep-th/0311179].
  %%CITATION = HEP-TH 0311179;%%


%\cite{Leblond:2005km}
\bibitem{Leblond:2005km}
  L.~Leblond,
  ``On the production of open strings from brane anti-brane annihilation,''
  arXiv:hep-th/0510261.
  %%CITATION = HEP-TH 0510261;%%

  
  \bibitem{Leblond:2004uc}
  L.~Leblond and S.-H.~H.~Tye,
  ``Stability of D1-strings inside a D3-brane,''
  JHEP {\bf 0403}, 055 (2004)
  [arXiv:hep-th/0402072]
  
  \bibitem{Polchinski:2005bg}
  J.~Polchinski,
  ``Open heterotic strings,''
  arXiv:hep-th/0510033.
  
  \bibitem{Blanco-Pillado:2005xx}
  J.~J.~Blanco-Pillado, G.~Dvali and M.~Redi,
  ``Cosmic D-strings as axionic D-term strings,''
  Phys.\ Rev.\ D {\bf 72}, 105002 (2005)
  [arXiv:hep-th/0505172].



%\cite{Barnaby:2004gg}
\bibitem{Barnaby:2004gg}
  N.~Barnaby, C.~P.~Burgess and J.~M.~Cline,
  ``Warped reheating in brane-antibrane inflation,''
  JCAP {\bf 0504}, 007 (2005)
  [arXiv:hep-th/0412040].
  %%CITATION = HEP-TH 0412040;%%


%\cite{Kofman:2005yz}
\bibitem{Kofman:2005yz}
  L.~Kofman and P.~Yi,
  ``Reheating the universe after string theory inflation,''
  Phys.\ Rev.\ D {\bf 72}, 106001 (2005)
  [arXiv:hep-th/0507257].
  %%CITATION = HEP-TH 0507257;%%


%\cite{Chialva:2005zy}
\bibitem{Chialva:2005zy}
  D.~Chialva, G.~Shiu and B.~Underwood,
  ``Warped reheating in multi-throat brane inflation,''
  arXiv:hep-th/0508229.
  %%CITATION = HEP-TH 0508229;%%


%\cite{Frey:2005jk}
\bibitem{Frey:2005jk}
  A.~R.~Frey, A.~Mazumdar and R.~Myers,
  ``Stringy effects during inflation and reheating,''
  arXiv:hep-th/0508139.
  %%CITATION = HEP-TH 0508139;%%


\bibitem{Klebanov:2000hb}
  I.~R.~Klebanov and M.~J.~Strassler,
  ``Supergravity and a confining gauge theory: Duality cascades and
  chiSB-resolution of naked singularities,''
  JHEP {\bf 0008}, 052 (2000)
  [arXiv:hep-th/0007191].
  
  
\bibitem{Krasnitz:2000ir}
  M.~Krasnitz,
  ``A two point function in a cascading N = 1 gauge theory from supergravity,''
  arXiv:hep-th/0011179.
  
  %\cite{Firouzjahi:2005qs}
\bibitem{Firouzjahi:2005qs}
  H.~Firouzjahi and S.-H.~H.~Tye,
  ``The shape of gravity in a warped deformed conifold,''
  arXiv:hep-th/0512076.
  %%CITATION = HEP-TH 0512076;%%


  %\cite{Dimopoulos:2001ui}
\bibitem{Dimopoulos:2001ui}
  S.~Dimopoulos, S.~Kachru, N.~Kaloper, A.~E.~Lawrence and E.~Silverstein,
  ``Small numbers from tunneling between brane throats,''
  Phys.\ Rev.\ D {\bf 64}, 121702 (2001)
  [arXiv:hep-th/0104239].
  %%CITATION = HEP-TH 0104239;%%

%\cite{Dimopoulos:2001qd}
\bibitem{Dimopoulos:2001qd}
  S.~Dimopoulos, S.~Kachru, N.~Kaloper, A.~E.~Lawrence and E.~Silverstein,
  ``Generating small numbers by tunneling in multi-throat  compactifications,''
  Int.\ J.\ Mod.\ Phys.\ A {\bf 19}, 2657 (2004)
  [arXiv:hep-th/0106128].
  %%CITATION = HEP-TH 0106128;%%
  
  \bibitem{Dienes:1998vg}
  K.~R.~Dienes, E.~Dudas and T.~Gherghetta,
  ``Grand unification at intermediate mass scales through extra dimensions,''
  Nucl.\ Phys.\ B {\bf 537}, 47 (1999)
  [arXiv:hep-ph/9806292].

 \bibitem{Cheng:2002ej}
  H.~C.~Cheng, J.~L.~Feng and K.~T.~Matchev,
  ``Kaluza-Klein dark matter,''
  Phys.\ Rev.\ Lett.\  {\bf 89}, 211301 (2002)
  [arXiv:hep-ph/0207125].  
  
    \bibitem{Servant:2002aq}
  G.~Servant and T.~M.~P.~Tait,
  ``Is the lightest Kaluza-Klein particle a viable dark matter candidate?,''
  Nucl.\ Phys.\ B {\bf 650}, 391 (2003)
  [arXiv:hep-ph/0206071].

%\cite{Verlinde:1999fy}
\bibitem{Verlinde:1999fy}
  H.~L.~Verlinde,
  ``Holography and compactification,''
  Nucl.\ Phys.\ B {\bf 580}, 264 (2000)
  [arXiv:hep-th/9906182].


%\cite{Starobinsky:1974}
\bibitem{Starobinsky:1974}
 A.~A.~Starobinsky and S.~M.~Churilov,
 Sov.\ Phys.\ JETP {\bf 38}, 1 (1974).


%\cite{Gibbons:1975kk}
\bibitem{Gibbons:1975kk}
  G.~W.~Gibbons,
  ``Vacuum Polarization And The Spontaneous Loss Of Charge By Black Holes,''
  Commun.\ Math.\ Phys.\  {\bf 44}, 245 (1975).
  %%CITATION = CMPHA,44,245;%%


%\cite{Page:1976df}
\bibitem{Page:1976df}
  D.~N.~Page,
  ``Particle Emission Rates From A Black Hole: Massless Particles From An
  %Uncharged, Nonrotating Hole,''
  Phys.\ Rev.\ D {\bf 13}, 198 (1976).
  %%CITATION = PHRVA,D13,198;%%


%\cite{Unruh:1976fm}
\bibitem{Unruh:1976fm}
  W.~G.~Unruh,
  ``Absorption Cross-Section Of Small Black Holes,''
  Phys.\ Rev.\ D {\bf 14}, 3251 (1976).
  %%CITATION = PHRVA,D14,3251;%%


%\cite{Dhar:1996vu}
\bibitem{Dhar:1996vu}
  A.~Dhar, G.~Mandal and S.~R.~Wadia,
  ``Absorption vs decay of black holes in string theory and T-symmetry,''
  Phys.\ Lett.\ B {\bf 388}, 51 (1996)
  [arXiv:hep-th/9605234].
  %%CITATION = HEP-TH 9605234;%%


%\cite{Das:1996we}
\bibitem{Das:1996we}
  S.~R.~Das, G.~W.~Gibbons and S.~D.~Mathur,
  ``Universality of low energy absorption cross sections for black holes,''
  Phys.\ Rev.\ Lett.\  {\bf 78}, 417 (1997)
  [arXiv:hep-th/9609052].
  %%CITATION = HEP-TH 9609052;%%


%\cite{Klebanov:1997kc}
\bibitem{Klebanov:1997kc}
  I.~R.~Klebanov,
  ``World-volume approach to absorption by non-dilatonic branes,''
  Nucl.\ Phys.\ B {\bf 496}, 231 (1997)
  [arXiv:hep-th/9702076].
  %%CITATION = HEP-TH 9702076;%%


%\cite{Merzbacher}
\bibitem{Merzbacher}
 E.~Merzbacher,
 Quantum Mechanics,
 John Wiley \& Sons, 1970.


%\cite{Iengo:2006gm}
\bibitem{Iengo:2006gm}
  R.~Iengo and J.~G.~Russo,
  ``Handbook on string decay,''
  arXiv:hep-th/0601072.
  %%CITATION = HEP-TH 0601072;%%
  %%Cited 0 times in SPIRES-HEP

  
  \bibitem{Garcia-Bellido:2001ky}
  J.~Garcia-Bellido, R.~Rabadan and F.~Zamora,
  ``Inflationary scenarios from branes at angles,''
  JHEP {\bf 0201}, 036 (2002)
  [arXiv:hep-th/0112147].
  
  \bibitem{Dasgupta:2002ew}
  K.~Dasgupta, C.~Herdeiro, S.~Hirano and R.~Kallosh,
  ``D3/D7 inflationary model and M-theory,''
  Phys.\ Rev.\ D {\bf 65}, 126002 (2002)
  [arXiv:hep-th/0203019].
  

%\cite{Buchel:2002wf}
\bibitem{Buchel:2002wf}
  A.~Buchel,
  ``Gauge / gravity correspondence in accelerating universe,''
  Phys.\ Rev.\ D {\bf 65}, 125015 (2002)
  [arXiv:hep-th/0203041].


%\cite{Giddings:2005ff}
\bibitem{Giddings:2005ff}
  S.~B.~Giddings and A.~Maharana,
  ``Dynamics of warped compactifications and the shape of the warped
  landscape,''
  arXiv:hep-th/0507158.
  %%CITATION = HEP-TH 0507158;%%


%\cite{Peccei:1977ur}
\bibitem{Peccei:1977ur}
  R.~D.~Peccei and H.~R.~Quinn,
  ``Constraints Imposed By CP Conservation In The Presence Of Instantons,''
  Phys.\ Rev.\ D {\bf 16}, 1791 (1977).
  %%CITATION = PHRVA,D16,1791;%%


%\cite{Wilczek:1977pj}
\bibitem{Wilczek:1977pj}
  F.~Wilczek,
  ``Problem Of Strong P And T Invariance In The Presence Of Instantons,''
  Phys.\ Rev.\ Lett.\  {\bf 40}, 279 (1978).
  %%CITATION = PRLTA,40,279;%%


%\cite{Weinberg:1977ma}
\bibitem{Weinberg:1977ma}
  S.~Weinberg,
  ``A New Light Boson?,''
  Phys.\ Rev.\ Lett.\  {\bf 40}, 223 (1978).
  %%CITATION = PRLTA,40,223;%%


%\cite{Goldberg:1983nd}
\bibitem{Goldberg:1983nd}
  H.~Goldberg,
  ``Constraint On The Photino Mass From Cosmology,''
  Phys.\ Rev.\ Lett.\  {\bf 50}, 1419 (1983).
  %%CITATION = PRLTA,50,1419;%%


%\cite{Tye:2005fn}
\bibitem{Tye:2005fn}
  S.~H.~Tye, I.~Wasserman and M.~Wyman,
  ``Scaling of multi-tension cosmic superstring networks,''
  Phys.\ Rev.\ D {\bf 71}, 103508 (2005)
  [Erratum-ibid.\ D {\bf 71}, 129906 (2005)]
  [arXiv:astro-ph/0503506].
  %%CITATION = ASTRO-PH 0503506;%%


%\cite{Ellis:1983ew}
\bibitem{Ellis:1983ew}
  J.~R.~Ellis, J.~S.~Hagelin, D.~V.~Nanopoulos, K.~A.~Olive and M.~Srednicki,
  ``Supersymmetric Relics From The big bang,''
  Nucl.\ Phys.\ B {\bf 238}, 453 (1984).
  %%CITATION = NUPHA,B238,453;%%


%\cite{Griest:1989wd}
\bibitem{Griest:1989wd}
  K.~Griest and M.~Kamionkowski,
  ``Unitarity Limits On The Mass And Radius Of Dark Matter Particles,''
  Phys.\ Rev.\ Lett.\  {\bf 64}, 615 (1990).
  %%CITATION = PRLTA,64,615;%%


%\cite{Greisen:1966jv}
\bibitem{Greisen:1966jv}
  K.~Greisen,
  ``End To The Cosmic Ray Spectrum?,''
  Phys.\ Rev.\ Lett.\  {\bf 16}, 748 (1966).
  %%CITATION = PRLTA,16,748;%%


%\cite{Zatsepin:1966jv}
\bibitem{Zatsepin:1966jv}
  G.~T.~Zatsepin and V.~A.~Kuzmin,
  ``Upper Limit Of The Spectrum Of Cosmic Rays,''
  JETP Lett.\  {\bf 4}, 78 (1966)
  [Pisma Zh.\ Eksp.\ Teor.\ Fiz.\  {\bf 4}, 114 (1966)].
  %%CITATION = JTPLA,4,78;%%


%\cite{Feng:2003xh}
\bibitem{Feng:2003xh}
  J.~L.~Feng, A.~Rajaraman and F.~Takayama,
  ``Superweakly-interacting massive particles,''
  Phys.\ Rev.\ Lett.\  {\bf 91}, 011302 (2003)
  [arXiv:hep-ph/0302215].
  %%CITATION = HEP-PH 0302215;%%


%\cite{Chung:1998ua}
\bibitem{Chung:1998ua}
  D.~J.~H.~Chung, E.~W.~Kolb and A.~Riotto,
  ``Nonthermal supermassive dark matter,''
  Phys.\ Rev.\ Lett.\  {\bf 81}, 4048 (1998)
  [arXiv:hep-ph/9805473].
  %%CITATION = HEP-PH 9805473;%%


\bibitem{Wittman:2000tc}
  D.~M.~Wittman, J.~A.~Tyson, D.~Kirkman, I.~Dell'Antonio and G.~Bernstein,
  ``Detection of weak gravitational lensing distortions of distant galaxies by
  cosmic dark matter at large scales,''
  Nature {\bf 405}, 143 (2000)
  [arXiv:astro-ph/0003014].


%\cite{McAllister:2004gd}
\bibitem{McAllister:2004gd}
  L.~McAllister and I.~Mitra,
  ``Relativistic D-brane scattering is extremely inelastic,''
  JHEP {\bf 0502}, 019 (2005)
  [arXiv:hep-th/0408085].
  %%CITATION = HEP-TH 0408085;%%




\end{thebibliography}
\end{document}